\documentclass[onecolumn,usenatbib]{mnras}

\usepackage{newtxtext,newtxmath}

\usepackage[T1]{fontenc}

\DeclareRobustCommand{\VAN}[3]{#2}
\let\VANthebibliography\thebibliography
\def\thebibliography{\DeclareRobustCommand{\VAN}[3]{##3}\VANthebibliography}

\usepackage{graphicx}	
\usepackage{amsmath}	

\newcommand{\fracbrac}[2]{\left(\frac{#1}{#2}\right)}
\newcommand{\paren}[1]{\left({#1}\right)}
\newcommand{\avg}[1]{\left\langle{#1}\right\rangle}
\newcommand{\AU}{\mathrm{~AU}}
\newcommand{\pd}[2]{\frac{\partial #1}{\partial #2}}
\newcommand{\dd}[2]{\frac{d #1}{d #2}}

\title[Scattered Disk Dynamics]{Scattered Disk Dynamics: The Mapping Approach}

\author[Hadden \& Tremaine]{
Sam Hadden,$^{1}$\thanks{E-mail: hadden@cita.utoronto.ca}
Scott Tremaine,$^{1,2}$
\\
$^{1}$Canadian Institute for Theoretical Astrophysics, 60 St. George St., Toronto, ON M5S 3H8, Canada\\
$^{2}$School of Natural Sciences, Institute for Advanced Study, Princeton, NJ 08540, USA
}

\date{Accepted XXX. Received YYY; in original form ZZZ}

\pubyear{2023}

\begin{document}
\label{firstpage}
\pagerange{\pageref{firstpage}--\pageref{lastpage}}
\maketitle

\begin{abstract}
We derive, and discuss the properties of, a symplectic map for the dynamics of bodies on nearly parabolic orbits. The orbits are perturbed by a planet on a circular, coplanar orbit interior to the pericenter of the parabolic orbit. The map shows excellent agreement with direct numerical integrations and elucidates how the dynamics depends on perturber mass and pericenter distance. We also use the map to explore the onset of chaos, statistical descriptions of chaotic transport, and sticking in mean-motion resonances.  
We discuss implications of our mapping model for the dynamical evolution of the solar system's scattered disk and other highly eccentric trans-Neptunian objects.
\end{abstract}




\section{Introduction}
\label{sec:intro}
Small bodies beyond the orbit of Neptune are understood to be the remnants of a formerly much larger planetesimal population that was dramatically sculpted by dynamical interactions with the outer planets during the solar system's early evolution \citep[e.g.,][]{gladman_transneptunian_2021}. The orbits of these trans-Neptunian objects (TNOs) have been influenced to varying degrees by the subsequent billions of years of evolution due to  various secular and resonant  dynamical phenomena, the Galactic tide and kicks from passing stars, as well as chaos \citep[see review by][]{saillenfest_long-term_2020}. Understanding this long-term dynamical evolution is necessary for connecting the orbits of the TNOs we observe today to the solar system's formation and early evolution. Similar mechanisms drive the evolution of the debris disks found around many young stars \citep{Wyatt2020} and presumably govern the production of interstellar objects such as `Oumuamua.

Many TNOs are on highly eccentric orbits. The most significant population of highly eccentric TNOs is the ``scattered disk'', comprised of objects for which chaotic dynamics induced by Neptune scattering leads to quasi-random semi-major axis evolution. The population of ``detached'' TNOs resides on large semi-major axis, high-eccentricity orbits similar to those of the scattered population, but these are not actively experiencing chaotic semi-major axis evolution because their perihelion distances are large enough that they do not make close approaches to Neptune or the other planets.  How these bodies were placed on their current orbits, dynamically isolated from any significant influence by the solar system's giant planets, is a puzzle. Finally, Neptune's distant mean-motion resonances (MMRs) are inferred to host large small-body populations on high-eccentricity orbits \citep[$e \gtrsim 0.5$; ][]{pike_51_2015,volk_ossos_2018,crompvoets_ossos_2022}. It is not clear whether these populations are comprised entirely of scattered-disk objects experiencing transient periods of ``resonance sticking'' \citep[e.g.,][]{yu_trans-neptunian_2018} or if some fraction of the populations are permanent captures from Neptune's outward migration early in the solar system's history. The high-eccentricity scattered, detached, and resonant populations are sometimes referred to collectively as ``extreme TNOs''  or eTNOs \citep[e.g.,][]{de_la_fuente_marcos_extreme_2014}. Drawing inferences about the solar system's dynamical history from the populations of eTNOs requires understanding the dynamics of highly eccentric orbits subject to planetary perturbations.

Analytic methods are desirable as a supplement to direct numerical simulations of eTNO dynamical evolution. Unfortunately, classical disturbing function expansions are generally of little use when considering the dynamics of the large eccentricities typical of eTNOs (though see \citealt{batygin_stability_2021}). To overcome these shortcomings, numerous previous studies have turned to mapping approaches to model the behaviour of highly eccentric orbits \citep[e.g.,][]{petrosky_chaos_1986,petrosky_area-preserving_1988,chirikov_chaotic_1989,malyshkin_keplerian_1999,pan_generalization_2004,shevchenko_kepler_2011,hadden_chaotic_2018}. These mapping models approximate particles' orbital evolution by considering the effects of successive impulsive energy ``kicks'' experienced by the particles during their pericenter passage. Mapping models thus provide a computationally efficient alternative to direct $N$-body simulations, which require time steps short enough to resolve Neptune's orbital period. 

In addition to the benefits of computational efficiency, the mapping approach lends itself to the application of theoretical results on planar area-preserving maps. There exists a rich theory of chaos and transport in planar symplectic maps \citep[see reviews by][]{mackay_transport_1984, meiss_symplectic_1992,meiss_thirty_2015}. This body of theory can be applied, for example, to understand when large-scale chaos is expected to occur, to describe how chaotic trajectories diffuse, and to explicate the role of ``resonance sticking'' in producing power-law escape time distributions from regions of phase space. All of these aspects of Hamiltonian dynamics are relevant to understanding the long-term evolution of scattered-disk objects and other eTNOs.

In this paper, we develop a mapping approach to the dynamics of highly eccentric orbits subject to an inner, planetary-mass perturber. We focus on the dynamics of eTNOs perturbed by Neptune, though the results can readily be generalized to other contexts.  
 This mapping provides an analytic model for the description of the mean-motion resonance dynamics of highly eccentric orbits. It also allows us to derive a statistical description of the evolution of scattered bodies using a Fokker--Planck approach. 

 We restrict ourselves to a solar system containing a single giant planet on a circular orbit (usually taken to have Neptune's mass and semi-major axis) and to bodies orbiting in the same plane as the planet. The latter restriction avoids complications such as Kozai-Lidov oscillations and should be a reasonable approximation for objects in the scattered and detached disks, which typically have only modest inclinations. 

 Our paper is structured as follows. In Section \ref{sec:the_map}, we introduce our iterative map equations and also derive a ``local'' approximation to those equations, valid when particles' fractional variations in orbital energy are small. 
 In Section \ref{sec:resonances_and_chaos}, we use our map equations to derive an integrable Hamiltonian model for the dynamics of mean-motion resonances (MMRs) of arbitrary order in the high-eccentricity regime. We also apply our MMR model to predict the boundary between chaotic and regular motion as a function of both pericenter distance and semi-major axis.
 In Section \ref{sec:diffusion}, we present a statistical description of the evolution of chaotic orbits and derive a Fokker--Planck equation governing the evolution of scattered bodies' orbital energies. Section \ref{sec:summary_and_conclusions} summarizes our conclusions.

\section{The Map}
\label{sec:the_map}
We approximate the dynamical evolution of test particles on  highly eccentric orbits using a twist map. We consider a test particle with semi-major axis $a$ and pericenter distance $q$ on a highly eccentric orbit about a star of mass $M_*$. The test particle is subject to perturbations from a planet of mass $m_p \ll M_*$ on a circular orbit of radius $a_p$; we assume that $a\gg a_p$ and $q\sim a_p$. The test particle is assumed to orbit in the same plane as the planet. We define the dimensionless quantities $\mu = m_p/M_*$, $x = a_p/a$ and $\beta = a_p/q$; we refer to $x$ as the ``energy'' since both $x$ and $E=-\frac{1}{2}GM_*/a$ are proportional to the inverse semi-major axis.  

Conservation of the Jacobi constant implies that, far from pericenter, the particle's Tisserand parameter, $T \propto \frac{1}{2}x+\beta^{-1/2}(2-x/\beta)^{1/2}$, is fixed. Since we are considering highly eccentric orbits with $a\gg a_p$ and $q\sim a_p$, a constant Tisserand parameter implies that during a pericenter passage the changes in  pericenter, $q$, and energy $x$ are related by $|\delta q/q|\sim  |\delta x| \ll 1$. Thus we may treat the particle's pericenter distance, $q$, as fixed.

Each time the test particle passes through pericenter, its interactions with the perturber result in an  energy ``kick'', moving it onto a new orbit with semi-major axis $a'$ and energy $x'=a_p/a'$.  As we show in Appendix \ref{appendix:kick_derivation}, this kick can be described by a function,  $F_\beta(\theta)$, such that $x' =x -2\mu \partial F_\beta(\theta)/\partial\theta$ where  $\theta$ is the angular separation between the test particle and perturber when the test particle is at pericenter. The test particle  returns to pericenter after a time $T'$ where $2\pi/T'=n_p{x'}^{3/2}$  and $n_p$ is the mean motion of the planet.  At this return, the angular separation between the test particle and perturber is $\theta' = \theta + n_pT'$. Therefore, we have the following iterative map describing the evolution of the test-particle orbit at successive pericenter passages:
 \begin{align}
     x' &= x - 2\mu \pd{F_\beta(\theta)}{\theta}\label{eq:map_full1} \\
    \theta' &= \theta + \frac{2\pi}{x'^{3/2}}~.\label{eq:map_full2}
 \end{align}
Equations \eqref{eq:map_full1} and \eqref{eq:map_full2} constitute a two-dimensional area-preserving twist map depending on two parameters, $\mu$ and $\beta$.

Previous studies have approximated the kick function, $\partial{F_\beta}/\partial{\theta}$, appearing in Equation \eqref{eq:map_full1} by means of series expansions \citep[e.g.,][]{petrosky_area-preserving_1988}, the interpolation of numerical integrals \citep[e.g.][]{pan_generalization_2004,hadden_chaotic_2018}, or fits to direct $N$-body integrations \citep{malyshkin_keplerian_1999}. We present a novel derivation of the Fourier series expansion of $F_\beta(\theta)$ in Appendix \ref{appendix:kick_derivation}. The advantage of this Fourier series representation of the kick function is that it enables us to predict the onset of dynamical chaos and chaotic diffusion rates as a function of pericenter distance (\S \ref{sec:chaos}). In particular, we show that,  for non-crossing orbits ($\beta < 1$), $F_\beta(\theta)$ can be written as the cosine series
\begin{equation}
    F_\beta(\theta) = \sum_{k=0}^\infty C_k(\beta) \cos(k\theta),
    \label{eq:F_cosine_series}
\end{equation}
where exact expressions for the coefficients $C_k(\beta)$ are given by Equation \eqref{eq:Ck_defn} in Appendix \ref{appendix:kick_derivation}. 

Numerical implementations of the mapping in Equations \eqref{eq:map_full1} and \eqref{eq:map_full2} require truncating the sum over index $k$ in Equation \eqref{eq:F_cosine_series} at a finite value. Generally, this is not a serious limitation because the coefficients decay exponentially with increasing $k$. Moreover, in Appendix \ref{appendix:kick_derivation} we show that they admit an asymptotic approximation $C_k(\beta)\approx A(\beta)e^{- k \lambda(\beta) }/k$ for constants $\lambda$ and $A$ that depend only on $\beta$ (eq.\ \ref{eq:asymp}). We take advantage of this asymptotic behaviour in our numerical implementation of the map by using the identity $\sum_{k=1}^\infty {k}^{-1}{e^{- k \lambda}}\cos(k\theta) = \frac{1}{2}\lambda -\frac{1}{2} \log [2 (\cosh \lambda  - \cos \theta)]$ to write 
\begin{equation}
    F_\beta(\theta) = -\frac{1}{2}A(\beta)
    \log\left[
    2(\cosh \lambda(\beta) - \cos\theta )
    \right] + \sum_{k=1}^{k_\mathrm{max}}\left
    [C_k(\beta) 
    - A(\beta)\frac{e^{- k \lambda(\beta)}}{k}
    \right]\cos(k\theta) + o(k_\mathrm{max}e^{-\lambda(\beta)k_\mathrm{max}}),
    \label{eq:F_asym}
\end{equation}
where we have dropped an unimportant constant term, $C_0(\beta) + \frac{1}{2}A(\beta)\lambda(\beta)$. We choose $k_\mathrm{max}$  to achieve a $5\%$ or better fractional error in the amplitudes ${C_k}$. The asymptotic approximation  $C_k(\beta)\approx A(\beta)e^{- k \lambda(\beta) }/k$, derived in Appendix \ref{appendix:kick_derivation}, is valid for $\beta < 8/9$, which corresponds to the limiting pericenter distance, as $x\rightarrow 0$, reached by particles that are initially on circular orbits co-orbital with the perturber. While our focus in this paper is on non-crossing orbits, we note that Equation \eqref{eq:F_asym} suggests that the functional form of the kick function for particles on crossing orbits ($\beta>1$) can be obtained by replacing the $\cosh \lambda$ term in Equation \eqref{eq:F_asym} with $\cos\theta_\mathrm{int}$ where $\theta_\mathrm{int}(\beta)$ is the orbital phase at which the particle's orbit intersects the planet's (see Eq.\ \ref{eq:theta_intersect}). Fitting the overall amplitude of the kick function derived in this manner may provide a simpler approximation of $N$-body results than the piecewise polynomial interpolation used by \citet{malyshkin_keplerian_1999} to represent the kick function of crossing orbits.

\subsection{Local approximation of map}
\label{sec:local_map}
Equations \eqref{eq:map_full1} and \eqref{eq:map_full2} provide an efficient yet accurate means of modeling the dynamics of comet-like orbits. Valuable insight into these dynamics is obtained by approximating the map locally in the vicinity of an $N$:1 mean-motion resonance, that is, when the planet orbits $N$ times for each orbit of the test particle. In order to do so, we will define $x_0 = N^{-2/3}$ as the nominal energy of the resonance and assume that variations about this nominal value, $\delta x$, remain small ($|\delta x / x_0|\ll 1$).  Writing $x = x_0\left( 1 -  \frac{2}{3N}w\right)$, the map can be approximated locally as 
\begin{align}
    w' &= w + 3 \mu N^{5/3}\pd{F_\beta(\theta)}{\theta} 
    \label{eq:map_local1}
    \\
    \theta' &= \theta  + 2\pi w'.    
    \label{eq:map_local2}
\end{align}
in which we have omitted $\mathcal{O}(w^2)$ terms in the second equation, under the assumption that  $|\delta x| / x_0\ll 1$. 
These equations again constitute a two-dimensional area-preserving twist map, with the dependence on both perturber mass and resonance index, $N$, entering through a single parameter, 
\begin{equation}
    \epsilon = 3\mu N^{5/3} = 3\mu (a_0/a_p)^{5/2}.
    \label{eq:epsdef}
\end{equation}

The map variable $w$ in Equations \eqref{eq:map_local1} and \eqref{eq:map_local2} has the following interpretation:  when $\epsilon=0$ the angle $\theta$ advances by exactly $2\pi(w+N)$ with each pericenter passage of the test particle. This means that at integer values, i.e., $w=m$ with integer $m$, the perturber completes $N+m$ orbits before the test particle returns to pericenter, corresponding to an $N+m:1$ mean-motion resonance (MMR). More generally, $w = p/q$ with $p$ and $q$ integers corresponds to $(Nq+p):q$ MMRs. We will refer to such a resonance as a $q$th order MMR. This differs from the standard nomenclature of celestial mechanics but, as noted by \citet{pan_generalization_2004}, is more appropriate when considering the dynamics of highly eccentric orbits with large semimajor axes.

The dynamics of the local map, Equations \eqref{eq:map_local1} and \eqref{eq:map_local2}, are unaltered by adding or subtracting integer values to the variable $w$ so that we can consider the dynamics by taking the value of $w$ modulo 1.\footnote{Note, however, when measuring $w$ modulo 1 in numerical simulations, it is not possible to determine whether $w$ can diffuse indefinitely or is bounded by KAM surfaces.} This invariance with respect to the addition of integer values to $w$ reflects a repeating pattern of MMRs between successive $N$:1 MMRs. These resonances occur at rational values of $w$ between 0 and 1.  In particular, all of the resonances up to a maximum order, $q$, occur at values of $w$ given by the Farey sequence, $F_q$, i.e., the sequence of reduced fractions between 0 and 1 having denominators less than or equal to $q$.
This repeating pattern of resonances is analogous to the repeating pattern of resonances between adjacent $j:j-1$ MMRs that are dynamically dominant for closely-spaced planet pairs as described in \cite{hadden_criterion_2018}.

\begin{figure}
    \centering
    \includegraphics[width=0.8\textwidth]{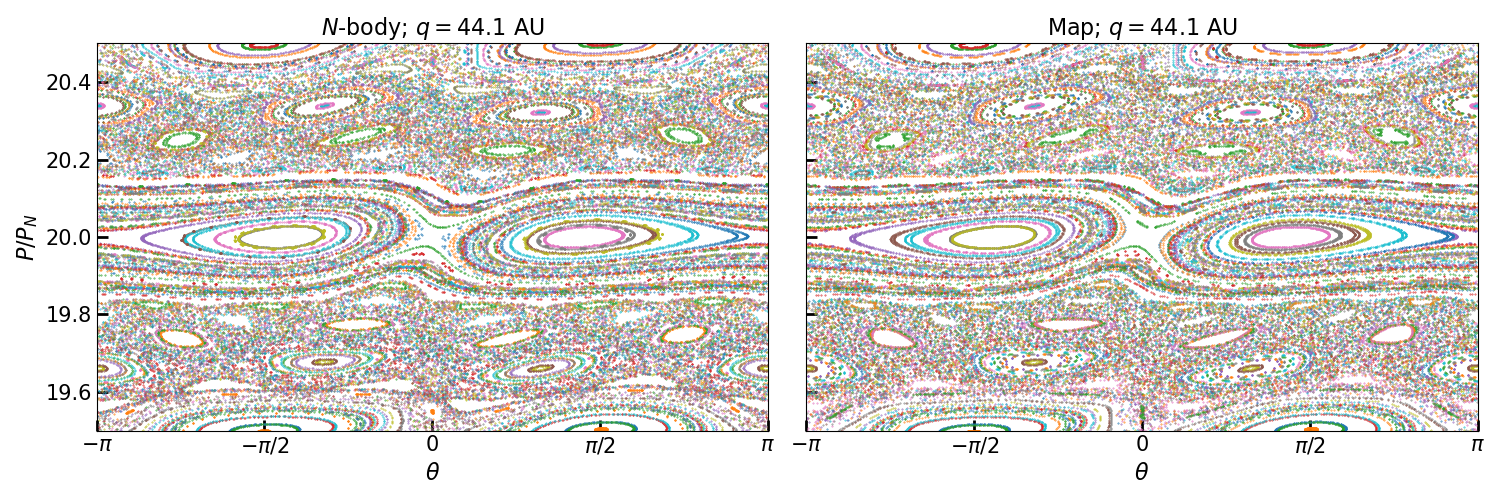}
    \caption{Comparison of the local map equations, Equations \eqref{eq:map_local1} and \eqref{eq:map_local2}, and direct $N$-body simulation. The map parameters are $\mu=5.15\times10^{-5}$, $N=20$, and $\beta=0.68$, corresponding to a perihelion distance $q= 44.1\AU$ for a planet at Neptune's semi-major axis. The $N$-body simulations follow particles near the 20:1 MMR with a perturber having Neptune's mass and semi-major axis. The vertical axis is the ratio of the orbital period to Neptune's orbital period.} 
    \label{fig:map_compare}
\end{figure}

Figure \ref{fig:map_compare} shows a comparison between direct $N$-body simulations and the local approximation of the map given by Equations \eqref{eq:map_local1} and \eqref{eq:map_local2}.  $N$-body integrations were done with the \texttt{ias15} integrator \citep{rein_ias15_2015} of the REBOUND code \citep{rein_rebound_2012}. The initial conditions are chosen near the orbit of TNO 148209 (2000 CR105) with a perihelion distance of $q= 44.1 \AU$ and near the 20:1 MMR with Neptune, which we model as a circular and coplanar perturber at semimajor axis $a_p=30\AU$. Figure \ref{fig:map_compare} can be compared with the surfaces of section shown in  Figure 1 of \cite{volk_orbital_2022}. The $N$-body surface of section in Figure \ref{fig:map_compare} is constructed from a set of initial conditions with fixed Jacobi constant.  Section points are generated by recording the test particle's osculating orbital period at aphelion and then the angular separation between the test particle and the perturber at its subsequent perihelion passage. 

\section{Resonances and the Onset of Chaos}
\label{sec:resonances_and_chaos}
\subsection{Dynamics of  Resonances}
\label{sec:resonances}
The twist map in Equations \eqref{eq:map_local1} and \eqref{eq:map_local2} can be derived 
from equations of motion 
\begin{align}
    \dd{\theta}{M} &= w + N \label{eq:theta_dot_local}\\
    \dd{w}{M} &= \epsilon\delta_{2\pi}(M)\pd{F_\beta(\theta)}{\theta} 
    \label{eq:w_dot}
\end{align}
where the independent variable, $M$, is the particle's mean anomaly, and 
\begin{equation*}
    \delta_{2\pi}(M) = \frac{1}{2\pi}\Big[1 + 2\sum_{j=1}^{\infty}\cos(jM)\Big]
\end{equation*}
is a $2\pi$-periodic delta function. Equations \eqref{eq:theta_dot_local}--\eqref{eq:w_dot} can be derived from the ``Hamiltonian'' function 
\begin{equation}
    \mathcal{F}(\theta,w,M;\beta)  = \frac{(w+N)^2}{2}
    - \frac{\epsilon}{2\pi}\sum_{j=-\infty}^\infty \sum_{k=1}^\infty C_{k}(\beta)
    \cos(jM + k\theta),
    \label{eq:Kamiltonian}
\end{equation}
where $(\theta,w)$  are a canonically conjugate coordinate-momentum pair and the particle's mean anomaly, $M$, plays the role of the independent ``time'' variable.

Let $w_0 = (J/K) - N$ denote the nominal value of $w$ at the $J$:$K$ MMR. Introducing new canonical variables $p = w - w_0$ and $\psi = \theta - (J/K)M$, dropping an unimportant constant, and only retaining resonant harmonics, the transformed Hamiltonian is  
\begin{equation}
 \mathcal{F}_\mathrm{res}(\psi,p) = \frac{1}{2}p^2 - \frac{\epsilon}{2\pi}\sum_{n=1}^\infty C_{nK}(\beta)\cos(nK\psi)~.\label{eq:res_model_local}
\end{equation}
Note that in the case of  first-order resonances (i.e., $K=1$), the sum over $n$ in Equation \eqref{eq:res_model_local} is simply $F_\beta(\psi)$.

\begin{figure}
     \centering
     \includegraphics[width=0.75\textwidth]{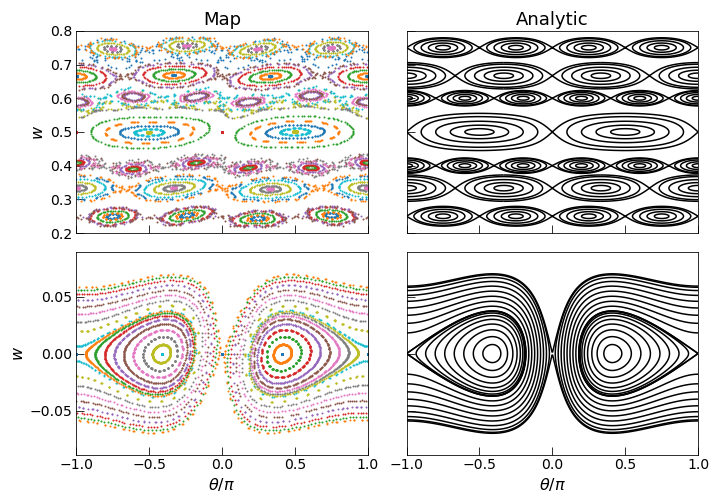}
     \caption{
     Comparison of resonant trajectories computed with the mapping to level curves of the analytic resonance Hamiltonian, Equation \eqref{eq:res_model_local}, for resonances up to order $K=5$. 
     The value of the analytic Hamiltonian is obtained as a function of $\theta$ and $w$ by substituting $(\psi,p)=(\theta,w-J/K)$ in Equation \eqref{eq:res_model_local}.
     The parameters of the map are $\epsilon = 0.005$ and $\beta=0.75$. For the case of particles subject to perturbations from Neptune, these parameters correspond to orbits with pericenter distances of $q\approx 40$ AU near the $8:1$ MMR.
     }
     \label{fig:map_vs_analytic}
 \end{figure}

Figure \ref{fig:map_vs_analytic} compares trajectories computed directly via the mapping, Equations \eqref{eq:map_local1} and \eqref{eq:map_local2}, to level curves of Hamiltonians for various resonances given by Equation \eqref{eq:res_model_local}. The phase space topology of $N$:$1$ MMRs, shown in the bottom row of Figure \ref{fig:map_vs_analytic}, is distinct from that of the higher-order MMRs shown in the top row of Figure \ref{fig:map_vs_analytic}. First-order  $N$:$1$ MMRs possess two unstable fixed points, at $(\psi,p)=(0,0)$ and $(\psi,p)=(\pi,0)$, along with two stable ``asymmetric'' libration islands situated at $p=0$ and  $\psi = \pm \psi_\mathrm{asym}(\beta)$ where $\psi_\mathrm{asym}$ satisfies  $\left[\pd{}{\theta}F_\beta(\theta)\right]_{\theta=\psi_\mathrm{asym}(\beta)} = 0$ and 
$\left[\pd{^2}{\theta^2}F_\beta(\theta)\right]_{\theta=\psi_\mathrm{asym}(\beta)}>0$.
The value of $w$ along the outer separatrices of first-order MMRs is given, as a function of $\psi$, by 
\begin{equation}
    w_\mathrm{sx}(\psi) = w_0 \pm 
    \sqrt{
        \frac{\epsilon}{\pi}
            \left[F_\beta(0) - F_\beta(\psi)\right]
        }~.
\end{equation} 

In contrast with $N$:1 MMRs, higher-order MMRs exhibit simpler phase space topologies, with each  $K^\mathrm{th}$ order MMR possessing a chain of $K$  pendulum-like libration islands. Since typically $C_K \gg C_{nK}$ for $n>1$, the dynamics of these MMRs can be approximated by retaining only the lowest harmonic from the sum appearing in Equation \eqref{eq:res_model_local}.  With this approximation $\mathcal{F}_\mathrm{res}$ becomes the Hamiltonian of a pendulum and we obtain a resonance half-width, measuring the distance from the elliptic fixed point to the separatrix, of $\Delta w_K =  2\sqrt{\epsilon |C_K(\beta)| /(2\pi)}$ for an $N$:$K$ MMR. In terms of semi-major axis, the full width of a $K^\mathrm{th}$ order MMR is approximately 
\begin{equation}
\fracbrac{\Delta a}{a}_K  \approx
\begin{cases}
4\fracbrac{\displaystyle a_p}{\displaystyle a}^{1/4}\sqrt{\frac{\displaystyle \mu}{\displaystyle 3\pi}}\sqrt{F_\beta(0) - F_\beta(\pi)}    &~ K = 1, 
\\[10 pt]
4\fracbrac{\displaystyle a_p}{\displaystyle a}^{1/4}\sqrt{\frac{\displaystyle 2\mu }{\displaystyle 3\pi}|C_K(\beta)|}    &~ K>1 ,
\end{cases}
\label{eq:res_width_da}
\end{equation}
in which we have approximated the width of a first-order MMR by its width at $\theta = \pi$. Note that the maximal width of first-order MMRs actually occurs at $\theta = \psi_\mathrm{asym}(\beta)$, but this width differs little from the width at $\psi=\pi$, which yields a slightly simpler expression in Equation \eqref{eq:res_width_da}.

\subsection{Onset of Chaos and Destruction of KAM Barriers}
\label{sec:chaos}
To predict the onset of large-scale chaos, we use the resonance covering fraction criterion developed in \citet{hadden_criterion_2018}  \citep[see also][]{quillen_three-body_2011}. This criterion predicts that resonance overlap and chaos occurs when the  total width of all resonances in a given region of phase space exceeds the size of that region.  Specifically, we follow \citet{hadden_criterion_2018} and define a ``resonance optical depth'', $\tau_\mathrm{res}$, as the covering fraction of resonances in a unit interval of the map variable $w$, and predict that chaos occurs when $\tau_\mathrm{res}=1$. The sum of resonance widths in a unit interval of the map variable $w$ is given by  
\begin{equation}
\label{eq:tau_res}
    \tau_\mathrm{res} = 2\sqrt{\frac{6\mu}{\pi}}\fracbrac{a}{a_p}^{5/4}\left[
    \sqrt{\frac{1}{2}(F_\beta(0) - F_\beta(\pi))}
    +
    \sum_{k=2}^{\infty}\phi(k)\sqrt{C_k(a_p/q)}
    \right],
\end{equation}
where $\phi(k)$, Euler's totient function, counts the number of resonances of order $k$ falling in a unit interval of $w$ 
\citep[see][]{hadden_criterion_2018}.\footnote{
Numerical calculation of $\tau_\mathrm{res}$ requires truncating the sum over $k$ at some finite value, $k_\mathrm{max}$. 
We approximate the contribution of the remaining terms of the sum in Equation \eqref{eq:tau_res} as follows. 
First, we approximate $C_k(\beta) \approx A(\beta)e^{-k\lambda(\beta)}/k$ for $k>k_\mathrm{max}$ (eq.\ \ref{eq:asymp}).
For large $k$, we approximate $\phi(k) \approx 6k/\pi^2$ since
$\sum_{l=1}^k\phi(l)  =  \frac{3k^2}{\pi^2} + \mathcal{O}\left[k(\log k)^{2/3}(\log\log k)^{4/3}\right]$. Finally, we approximate the sum over the remaining terms as an integral using
$
\sum_{k=k_\mathrm{max}+1}^\infty
 \sqrt{k}e^{-\frac{\lambda}{2}k} 
 \approx 
 \int_{k_\mathrm{max}+1/2}^\infty dk 
 \sqrt{k}e^{-k\lambda/2} 
 = 
 \paren{2/\lambda}^{3/2} \Gamma \left[\frac{3}{2},\frac{1}{2} \lambda  \left(k_{\max }+\frac{1}{2}\right)\right]
$
where $\Gamma[\cdot,\cdot]$ is the incomplete Gamma function.
}

\begin{figure}
    \centering
    \includegraphics[width=0.8\textwidth]{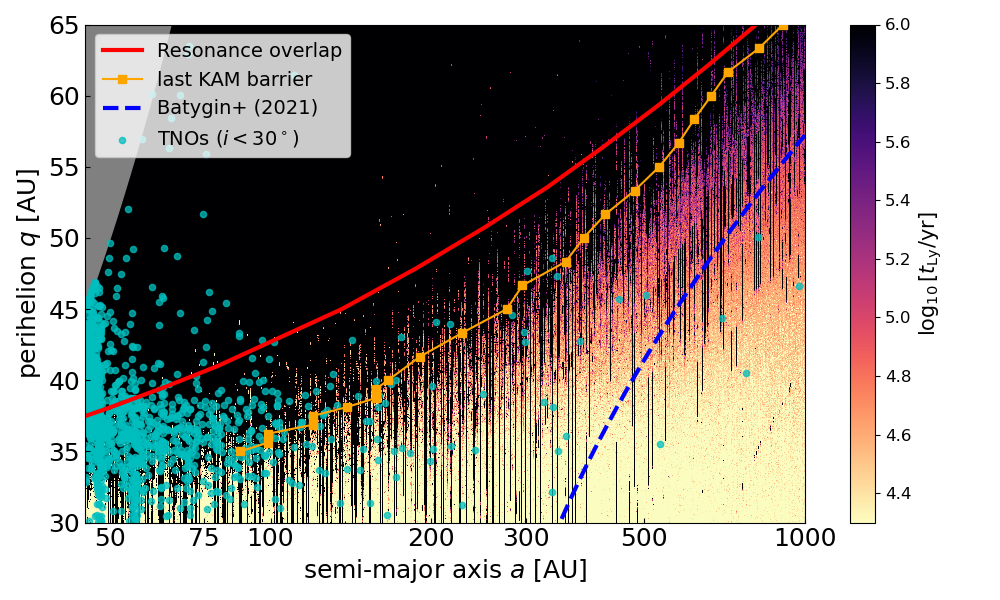}
    \caption{Dynamical map in semi-major axis, $a$, versus perihelion distance, $q$, for test particles perturbed by Neptune $(m/M_\odot = 5.15\times10^{-5}, a = 30$ AU). The color scale shows the value of the Lyapunov time computed on a grid of $N$-body simulations. The red line shows the analytic prediction of the resonance overlap criterion, setting $\tau_\mathrm{res}=1$ in Equation \eqref{eq:tau_res}. The blue line shows the analytic resonance overlap criterion of \citet{batygin_stability_2021}. 
    Observed low-inclination  TNOs ($i<30^\circ$) are plotted as cyan points.
    See text for additional details.}
    \label{fig:overlap_prediction}
\end{figure}

In Figure \ref{fig:overlap_prediction} we plot our prediction of the onset of chaos, where $\tau_\mathrm{res}=1$, on a plot of semi-major axis versus perihelion distance for particles subject to a Neptune-mass perturber on a circular orbit at $30$ AU. 
We compare our prediction to a grid of Lyapunov times computed via $N$-body simulations that indicate where chaotic behaviour occurs.
Each $N$-body simulation was integrated for a total of 50 test particle orbits 
using the \texttt{WHFast} integrator of the REBOUND code \citep{rein_whfast_2015}, an implementation of the Wisdom-Holman symplectic mapping algorithm \citep{wisdom_symplectic_1991} that includes variational equations. Variational equations were integrated along each orbit and used to compute  Lyapunov times, taken to be the inverse of the Lyapunov characteristic exponent computed via the \texttt{Simulation.calculate\_lyapunov} method in REBOUND.

We compare our prediction for resonance overlap and the onset of chaos to the resonance overlap boundary derived by \cite{batygin_stability_2021}, shown as a blue dashed curve in Figure \ref{fig:overlap_prediction}. 
The criterion derived by \cite{batygin_stability_2021} accounts only for the overlap of $N$:$1$ and $N$:$2$ MMRs and the disagreement between their prediction and the numerical results, especially for small $q$, underscores the increasing dynamical importance of high-order $N:k$ MMRs with $k>2$ as $\beta \rightarrow 1$.\footnote{
The formulae that \cite{batygin_stability_2021} derive for the widths of $N$:1 and $N$:2 MMRs also differ from those predicted by Equation \eqref{eq:res_width_da}. 
}  

The long-term evolution of chaotic trajectories will be qualitatively different depending on whether or not their motion is bounded by KAM tori, which serve as absolute barriers to transport in the phase space of our two-degree-of-freedom system.
We conduct numerical experiments with the map to predict the location of the last KAM barrier in phase space as a function of pericenter distance.
We adopt the following procedure to estimate the location of the last KAM curve:
\begin{enumerate}
    \item 
    We set a fixed pericenter distance and select an initial value $N$ to determine the parameter $\epsilon$ appearing in our local approximation of the map (Equation \ref{eq:epsdef}).
    \item 
    We initialize ten trajectories in the vicinity of the point $(\theta,w) = (0,1/2)$, corresponding to the unstable fixed point of the $2N+1$:2 MMR.
    \item The trajectories are iterated up to a maximum of $3\times 10^7 / N$ times (corresponding to a time of approximately $5\,$Gyr) or until any of the trajectories escapes the region $0<w<1$.
    \item If any trajectory escapes, we reduce $N$ by 1 and repeat the experiment. Otherwise, we take $a = a_p \times N^{2/3}$ as the semi-major axis of the last KAM curve.
\end{enumerate}
The computed locations of the last KAM barriers are plotted in Figure \ref{fig:overlap_prediction}.
For large $q$, the location of the last KAM barrier closely matches the boundary for the onset of chaos while at smaller pericenter distances there is an increasing gap between the onset of chaotic behaviour and the location of the last KAM curve, indicating a growing region of bounded chaos.

We plot the pericenter and semi-major axis distances of TNOs with 
$45\mathrm{~AU}<a<1000\mathrm{~AU}$ and $30\mathrm{~AU}<q<65\mathrm{~AU}$ obtained from a query of the Minor Planet Center (MPC) database.\footnote{\url{https://minorplanetcenter.net/db_search/}.
Note that the detached objects Sedna \citep[$q=76\AU~,~a=506\AU$;][]{brown_discovery_2004} and 2012 VP$_{113}$ \citep[$q=85\AU~,~a=262\AU$;][]{trujillo_sedna-like_2014} fall outside the  plotted pericenter range.
} 
Figure \ref{fig:overlap_prediction}  shows that essentially all TNOs with $a\gtrsim 100\AU$ are within the predicted resonance overlap region.
Furthermore, these objects are generally near or below the curve marking the last KAM barrier at a given pericenter distance and are therefore subject to large-scale migration in semi-major axis. (Additional dynamical effects arising from the other giant planets, Neptune's eccentricity, and inclination will generally render the KAM boundary plotted in Figure \ref{fig:overlap_prediction} a partial, rather than absolute, barrier to transport for objects slightly above the curve.)
There is an apparent lack of low perihelion distance objects 
($q\lesssim40\AU$) at large semi-major axes, despite a strong observational bias towards discovery of low perihelion orbits. We show below that the rate of chaotic diffusion exhibits a steep  (i.e., exponential) dependence on pericenter distance (Fig.\ \ref{fig:TD_vs_q}) and the relative excess of high-perihelion distance objects is likely attributable to the exponentially longer expected survival times at higher pericenter distance. 
Since quasi-circular orbits with $q\gtrsim 40\AU$ are far from the resonance-overlap region and cannot reach large semi-major axes via chaotic scattering with Neptune the high-$q$, large $a$ objects presumably owe their present orbits to torques from  stellar passages or the Galactic tide experienced while scattering to large $a$ from lower values of $q$ \citep{vokrouhlicky_origin_2019}, or perhaps to a rogue planet \citep{gladman_production_2006,silsbee_producing_2018}.

\section{Statistical Description of Ensemble Evolution}
\label{sec:diffusion}

Here we consider an ensemble of chaotic particle orbits and seek a statistical description of their evolution.  We begin in Section \ref{sec:diffusion:local_diffusion} with an analysis of chaotic diffusion in the local approximation of the map that was presented in Section \ref{sec:local_map}. We derive the local map's chaotic diffusion rate as a function of the parameters $\beta=a_p/q$ and $\epsilon$ (Equation \ref{eq:epsdef}). Then, in Section \ref{sec:diffusion:FP} we derive a Fokker--Planck description of the evolution of the energy of the scattering particles.  While modeling comet energy evolution via a diffusion equation is not new \citep[e.g.,][]{van_woerkom_origin_1948,yabushita_exact_1980}, past works have assumed that the rate of diffusion per pericenter passage is independent of energy. This approximation is valid  as $a\rightarrow\infty$ but, as we show in Section \ref{sec:diffusion:local_diffusion}, breaks down for semi-major axes  $a\lesssim 1000$ AU where the majority of known TNOs reside.  We also solve the Fokker--Planck equation with boundary conditions that more accurately capture the behaviour of scattered disk objects than those used in past works. We compare our Fokker--Planck description of orbital evolution to direct $N$-body simulations in Section \ref{sec:diffusion:nbody_compare}. Finally, in Section \ref{sec:sticking} we examine deviations from diffusive behaviour exhibited by the map that result from resonant sticking.

\subsection{Diffusion in the local approximation}
\label{sec:diffusion:local_diffusion}

For chaotic trajectories, the evolution of the map variable $w$, when viewed on sufficiently long time scales in regions of phase space where no KAM barriers are present, is well-approximated as a random walk and the mean square displacement of an initial ensemble of nearby trajectories will increase linearly in time. We define a  normalized diffusion coefficient for this random walk  (see Appendix \ref{sec:diffusion-force} for intermediate steps)
\begin{equation}
      \mathcal{\mathcal{D}(\epsilon,\beta)} :=
      \lim_{n\rightarrow\infty}
        \frac{\langle{(w_n-w_0)^2}\rangle}{\epsilon^2 n} 
   =
        \mathcal{D}_\mathrm{QL}(\beta) 
            + 
        2\sum_{k=1}^\infty \langle a_0a_k\rangle\label{eq:diff_coeff}~,
\end{equation}
where $\epsilon$ is the map parameter defined in Equation \eqref{eq:epsdef}, angle brackets indicate averages over initial values of $\theta$ and $w$, i.e.,  $\langle f(\theta,w)\rangle=(2\pi)^{-1}\int_{-\pi}^{\pi} d\theta\int_{-1/2}^{1/2}dw f(\theta,w)$,
$a_i = (w_{i+1}-w_i)/\epsilon$, and  $ \mathcal{D}_\mathrm{QL}(\beta) = \langle a_0^2\rangle=\frac{1}{2}\sum_{k=1}^{\infty}{k^2}C_k(\beta)^2$ is referred to as the ``quasi-linear'' approximation of the diffusion rate \citep[see, e.g.,][]{karney_effect_1982}.

The  terms $\langle a_0a_k \rangle$ capture correlations between kicks $k$ iterates apart and can be computed as follows: the functions $a_k(\theta_0,w_0)$ are periodic in both $\theta_0$ and $w_0$ and so can be written in terms of a Fourier expansion as  $a_k(\theta_0,w_0) = \sum_{l,m\in \mathbb{Z}}\hat{a}_{k}^{(l,m)}\exp[{i(2\pi l w_0 + m \theta_0)}]$. Additionally, it follows from the definition of the functions $a_k(\theta_0,w_0)$ that  $\hat{a}_0^{(l,m)}=\frac{1}{2}im\delta_{l0}C_{|m|}(\beta)$. Therefore, each correction term can be expressed as the following  sum over Fourier amplitudes:
\begin{equation}
    \langle a_0a_k\rangle = -\frac{i}{2}\sum_{m=1}^{\infty}mC_m(\beta)\hat{a}_{k}^{(0,m)} + c.c.~,
    \label{eq:diff_coeff_correction}
\end{equation}
where $c.c.$ denotes the complex conjugate of the preceding term.

\citet{cary_rigorously_1981} describe a procedure for expressing the correlation terms, $\langle a_0a_k\rangle$, analytically for a broad class of maps that includes our Equations \eqref{eq:map_local1} and \eqref{eq:map_local2}. Using their procedure, it is straightforward to show that $\langle a_0a_1 \rangle = 0$. However, the analytic expressions for higher-order correlations are unduly cumbersome due to the fact that the function $\partial_\theta F_\beta(\theta)$ appearing in Equation \eqref{eq:map_local1} has a non-zero Fourier amplitude at every harmonic. Therefore, we implement a numerical procedure to compute correlation terms $\langle a_0a_k \rangle$ up to some maximum order, $K$, and thereby obtain an estimate of the diffusion coefficient in Equation \eqref{eq:diff_coeff}, which we denote as
\begin{equation}
\mathcal{D}_{K}(\epsilon,\beta) = \mathcal{D}_\mathrm{QL}(\beta)  + 2\sum_{k=2}^K \langle a_0a_k\rangle~.\label{eq:diff_coeff_K}
\end{equation}
To calculate $\mathcal{D}_{K}(\epsilon,\beta)$, we compute $K$ iterates of the map on a grid of initial $(\theta_0,w_0)$ values, recording $a_{k}(\theta_0,w_0)$ for $k=1,\ldots,K$ at each grid point. Using the grid of computed $a_{k}(\theta_0,w_0)$ values, amplitudes $\hat{a}_{k}^{(l,m)}$ are computed by means of a fast Fourier transform (FFT) algorithm. The resulting amplitudes are then used to compute the sum in Equation \eqref{eq:diff_coeff_correction}.
These are in turn used to compute $\mathcal{D}_{K}(\epsilon,\beta)$ according to Equation \eqref{eq:diff_coeff}  with the sum over $k$ limited to $k\le K$.

\begin{figure}
    \centering
    \includegraphics[width=0.85\textwidth]{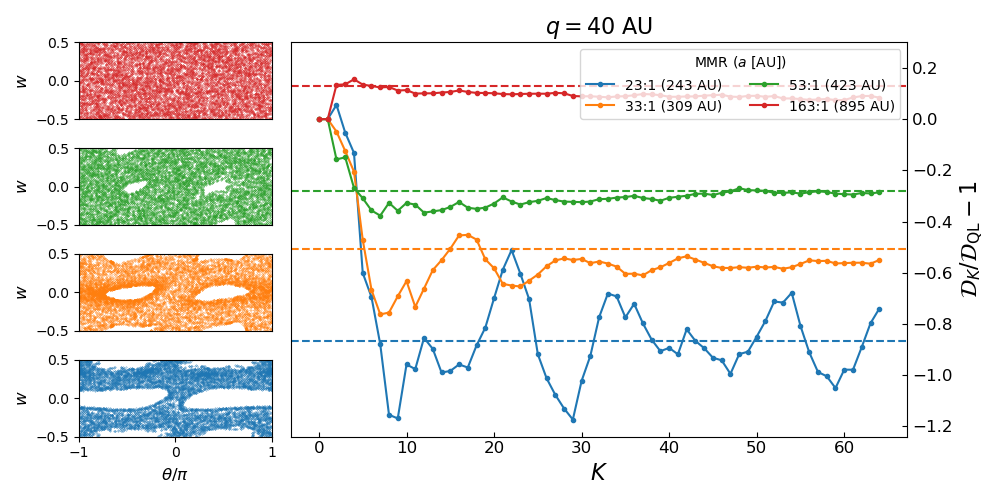}
    \caption{
    The left column shows trajectories of individual chaotic orbits of the local map, Equations \eqref{eq:map_local1} and \eqref{eq:map_local2}, in the vicinity of various $N:1$ MMRs of a Neptune-mass perturber ($\mu=5.15\times 10^{-5}$, $a=30~\mathrm{AU}$) for a pericenter distance of $q=40~\mathrm{AU}$ (i.e., $\beta = q/a = 0.75$).
    The right panel shows the fractional difference between the diffusion rate $\mathcal{D}_K(\epsilon,\beta)$ defined in Equation \eqref{eq:diff_coeff_K} and the quasi-linear diffusion estimate $\mathcal{D}_\mathrm{QL}$, as a function of the number, $K$, of correlation terms, $\langle a_0 a_k\rangle$, included in the calculation of the diffusion coefficient. Dashed horizontal lines show diffusion rates measured numerically from ensembles of map trajectories, as described in the text.}
    \label{fig:diff_coeff_vs_kmax}
\end{figure}

Figure \ref{fig:diff_coeff_vs_kmax} illustrates the behaviour of  $\mathcal{D}_{K}(\epsilon,\beta)$ as a function of $K$ for orbits with pericenter distance $q=40\AU$ subject to a Neptune-like perturber, at four different semi-major axes. The semi-major axes range from  $a\sim 240\AU$, just after the last KAM barrier has disappeared, to $a\sim 900\AU$, corresponding  to a regime of strong resonance overlap (see Figure \ref{fig:overlap_prediction}). As the semi-major axis decreases, regular trajectories associated with low-order MMRs occupy progressively more phase-space volume. 
At the 23:1 MMR ($a = 243\AU$), where regular resonant islands take up the largest fraction of the phase space, the diffusion coefficient estimates $\mathcal{D}_{K}(\epsilon,\beta)$ show large oscillations about the numerically measured value, even at large $K$. These oscillations arise from the contributions of the regular islands to the ensemble-averaged correlations, $\langle a_0a_k\rangle$, a generic feature of force correlations in maps with a mixed phase space \citep[e.g.,][]{karney_effect_1982}. We are interested in the diffusion rate of trajectories in the connected chaotic component of the map, which we assume to be well-approximated by the mean value, $\approx \frac{1}{K_\mathrm{max}}\sum_{K=0}^{K_\mathrm{max}}\mathcal{D}_{K}(\epsilon,\beta)$, for sufficiently large $K_\mathrm{max}$ such that the oscillating contributions to $\langle a_0a_k\rangle$ from regular islands are averaged out.

\begin{figure}
    \centering
    \includegraphics[width=0.85\textwidth]{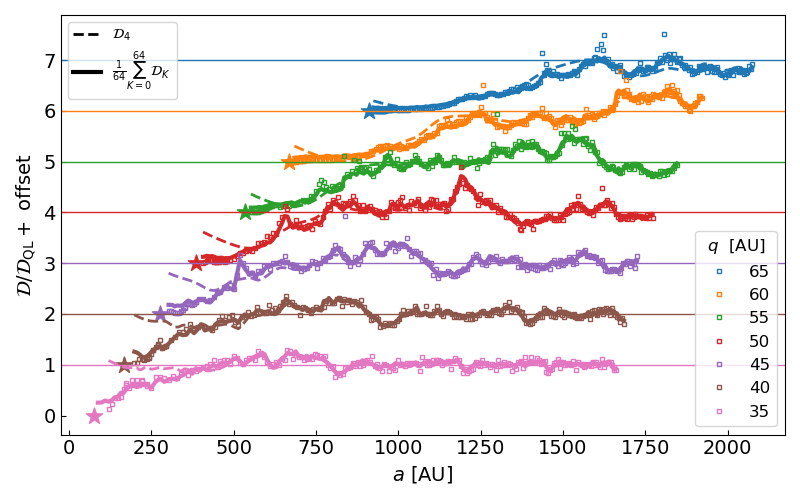}
    \caption{ Comparison of the numerically measured diffusion rate of a ball of initially close trajectories and our semi-analytic method for predicting diffusion rates using Equation \eqref{eq:diff_coeff_K} (see text for details).
    Diffusion rates normalised by the quasi-linear rate are plotted as a function of semi-major axis for different pericenter distances. An arbitrary offset is included for each pericenter distance to improve legibility.
    Numerically estimated values are plotted as open squares.  
    Lines show our semi-analytic estimates of the diffusion coefficient, computed as either $\mathcal{D}(\epsilon,\beta) \approx \frac{1}{64}\sum_{K=0}^{64}\mathcal{D}_K(\epsilon,\beta)$ (solid) or $\mathcal{D}(\epsilon,\beta) \approx \mathcal{D}_4(\epsilon,\beta)$ (dashed).
     Stars indicate the estimated semi-major axis of the last KAM curve for a given pericenter distance, derived according to the procedure described in Section \ref{sec:chaos}.
    }
    \label{fig:diffusion_compare}
\end{figure}

In Figure \ref{fig:diffusion_compare} we calculate mean diffusion coefficients over a range of pericenter and orbital distances, taking $K_\mathrm{max}=64$. We compare these {semi-analytic} diffusion coefficients to numerical measurements derived by evolving ensembles of chaotic trajectories. The numerical measurements are obtained by initializing 1000 trajectories  with $\theta_0 = 0$ and $w_0$ drawn from a Gaussian distribution centered on $w_0=1/2$ with dispersion $\sigma = 10^{-3}$ so that the initial ensemble resides near the unstable period-2 orbit associated with an $2N:2N+1$ MMR.
The trajectories of the ensemble are evolved for 5,000 iterations and  the  variance of the ensemble's $w$ values is recorded at each iteration. The diffusion coefficient, $\mathcal{D}(\epsilon,\beta)$, is then estimated from a linear fit of the variance versus iteration data. Both numerical and semi-analytic results show that the diffusion rate declines to zero as the location of the last KAM curve, indicated by stars for each pericenter distance in Figure \ref{fig:diffusion_compare}, is approached. We also plot the values of ${\mathcal{D}}_4(\epsilon,\beta)$ for the different pericenter distances, shown as  dashed lines in Figure \ref{fig:diffusion_compare}, to highlight that good agreement with numerical results can be obtained with relatively few low-order corrections to the quasi-linear diffusion rate. However, we caution that the good agreement between ${\mathcal{D}}_4(\epsilon,\beta)$ and the numerical results is deceiving at semi-major axes near the last KAM curve: there, the values ${\mathcal{D}}_K$  show large oscillations about the numerically-measured value for $K>4$, similar to the case of the 23:1 MMR in Figure \ref{fig:diff_coeff_vs_kmax}, so that the agreement is generally worse if more terms are added.  Nonetheless, we find empirically that the numerically determined diffusion rates are nearly always well-approximated by ${\mathcal{D}}_4(\epsilon,\beta)$ except very near the semi-major axis of the last KAM curve. Thus,  ${\mathcal{D}}_4(\epsilon,\beta)$ is useful as a relatively accurate estimate of the diffusion rate that can be calculated very efficiently.

\subsection{Formulation of Fokker--Planck Equation}
\label{sec:diffusion:FP}
We now use our results on the diffusion rates of chaotic trajectories in the local version of the mapping to obtain a statistical description of the chaotic evolution of test particles in the restricted three-body problem. Consider the evolution of an ensemble of test particles. Let $n(x,t)dx$ be the number density of particles with inverse semi-major axes in the range $[x,x+dx)$ at time $t$. We assume that the ensemble members are randomly distributed in orbital phase. Let $W(x'|x)dx'$ be the probability that a particle with inverse semi-major axis $x$ transitions to an orbit with inverse semi-major axis in the range $[x',x'+dx')$ in one orbital period $P_p$ of the perturber due to a gravitational interaction. Then 
\begin{equation}
    \pd{n(x,t)}{t}=\frac{1}{P_p}\int_{-\infty}^\infty \left[
    n(x-\delta,t)W(x|x-\delta) 
    -  
    n(x,t)W(x-\delta|x)
    \right]d\delta=
    \frac{1}{P_p}\int_{-\infty}^\infty 
    n(x-\delta,t)W(x|x-\delta) d\delta
    -  
   \frac{n(x,t)}{P_p}.
    \label{eq:maseter_eq}
\end{equation}
In order to derive the Fokker--Planck equation, the integrand in the right-hand side of Equation \eqref{eq:maseter_eq} is written as $n(x-\delta,t)W(x+\delta-\delta|x-\delta)$, then expanded as a Taylor series in $-\delta$ to obtain
\begin{equation}
\label{eq:maseter_eq_taylor}
    \pd{n(x,t)}{t}=
            \frac{1}{P_p}\sum_{k=1}^\infty\frac{1}{k!}
        \int_{-\infty}^\infty
        (-\delta)^k
         \pd{^k}{x^k}\left[n(x,t) W( x + \delta| x)\right]
        d\delta~.
\end{equation}
Truncating the expansion at second order\footnote{
\citet{pawula_approximation_1967} shows that if a so-called ``Kramers-Moyal expansion'' such as Equation \eqref{eq:maseter_eq_taylor} is approximated by a truncation of finite order, then only truncation at $k=1$ or $k=2$ leads to logically consistent approximate equations.
} and integrating by parts, we obtain 
\begin{equation}
    \pd{n(x,t)}{t} = -\pd{}{x}\left[B(x) n(x,t)\right] + \frac{1}{2}\pd{^2}{x^2}\left[D(x) n(x,t)\right],
    \label{eq:FP_eqn}
\end{equation}
where 
\begin{align}
    B(x) &= \frac{1}{P_p}\int_{-\infty}^{\infty} (x'-x)W(x'|x) dx'
    \\
    D(x) &= \frac{1}{P_p}\int_{-\infty}^{\infty} (x'-x)^2 W(x'|x) dx'~
    \label{eq:bddef}
\end{align}
are referred to as the drift and diffusion coefficients, respectively.

In Section \ref{sec:diffusion:local_diffusion}, we showed that the mean square kick in $w=\frac{3}{2}N(1-x/x_0)$ \emph{per perihelion passage} is given by $\epsilon^2\mathcal{D}(\epsilon,\beta)$. Noting that $\epsilon = 3\mu N^{5/3}$, $x_0=N^{-2/3}$, and that a particle's orbital period is equal to $P_p x^{-3/2}$, this translates to a mean square kick in $x$ \emph{per perturber orbital period} of  $D(x) = 4P_p^{-1}\mu^2x^{3/2}\mathcal{D}(3\mu x^{-5/2},\beta)$, which is precisely the diffusion coefficient appearing in Equation \eqref{eq:FP_eqn}.

Deriving the drift coefficient, $B(x)$, nominally requires working out the expected kick to second order in $\mu$, whereas our map has only approximated the kick function to first order in $\mu$. However, we can take advantage of the fact that the underlying equations governing the particles' dynamics are Hamiltonian to derive a relationship between $B(x)$ and $D(x)$. We provide such a derivation, working directly from Hamilton's equations, in Appendix \ref{sec:second_order_direct_derivation}. Here we provide a shorter alternative derivation based on the concept of detailed balance.

\begin{figure}
    \centering
    \includegraphics[width=0.5\textwidth]{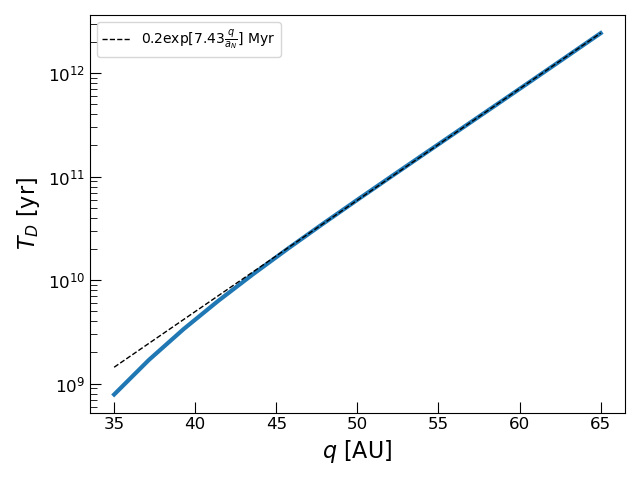}
    \caption{Diffusion timescale, $T_D$, versus perihelion distance, along with the approximate analytic expression, Equation \eqref{eq:TD_scaling}.}
    \label{fig:TD_vs_q}
\end{figure}

Since the Hamiltonian flow must preserve phase space volume, a distribution that is uniform in both the action $\Lambda \propto 1/\sqrt{x}$ and its associated angle variable $\lambda$ is stationary. So $n(x,t)dx \propto x^{-3/2}dx$ is a stationary distribution. Equation \eqref{eq:FP_eqn} says that the current in $x$ is $B(x)n(x,t)-\frac{1}{2}\pd{}{x}D(x)n(x,t)$.  Since detailed balance means that for an equilibrium stationary distribution there  are no net probability currents, we obtain 
\begin{equation}
    B(x)=\frac{x^{3/2}}{2}\dd{}{x}\left[x^{-3/2}D(x)\right]~.
    \label{eq:fluc-diss}
\end{equation}
The resulting Fokker--Planck equation can then be written as 
\begin{equation}
    \pd{n(x,t)}{t} = \frac{1}{2T_D}\pd{}{x}\left[
    \frac{\mathcal{D}(\epsilon(x),\beta)}{\mathcal{D}_\mathrm{QL}(\beta)}
    \pd{}{x}
    \paren{x^{3/2}n(x,t)}
    \right]~,
    \label{eq:FP_eqn_simpler}
\end{equation}
where $T_D = \frac{P_p}{4\mu^2 \mathcal{D}_\mathrm{QL}}$ and $\epsilon(x) =3\mu x^{-5/2}$. Figure \ref{fig:TD_vs_q} shows the diffusion timescale, $T_D$, as a function of  pericenter distance for a Neptune-like perturber. The figure illustrates that the diffusion timescale is a steep function of pericenter distance, and for $q\gtrsim 40\mathrm{~AU}$, is well approximated by the exponential relationship
\begin{equation}
    T_D \approx 
    0.2\mathrm{~Myr}\times \fracbrac{m_p}{ 17 M_\oplus}^{-2}
    \fracbrac{a_p}{30\mathrm{~AU}}^{3/2}
    \fracbrac{M_*}{M_\odot}^{3/2}
    \exp\left[7.4\fracbrac{q}{a_p}\right]~,
    \label{eq:TD_scaling}
\end{equation}
where we have normalized the scaling relationship Equation \eqref{eq:TD_scaling} such that the nominal parameters correspond to Neptune's mass and orbital distance.

When the diffusion coefficient in Equation \eqref{eq:FP_eqn_simpler} is approximated by the quasi-linear value, i.e., when  the $x$ dependence of ${\mathcal{D}(\epsilon(x),\beta)}$ is ignored, then the Fokker--Planck equation reduces to the diffusion equation modeled by \citet{yabushita_exact_1980}.   He derives the analytic solution    
\begin{equation}
    \label{eq:yabushita_soln}
    n_\mathrm{QL}(x,t;x_0) = \frac{4 \sqrt{x_0}}{(t/T_D)} \frac{1}{x}I_2\left[\frac{16 }{(t/T_D)}(xx_0)^{1/4}\right]\exp\left[-\frac{8}{(t/T_D)}\paren{\sqrt{x_0} + \sqrt{x}}\right],
\end{equation}
 where $I_1$ is a modified Bessel function, as the Green's function for Equation \eqref{eq:FP_eqn_simpler} assuming initial conditions $n(x,0;x_0) = \delta(x-x_0)$ and subject to the boundary conditions $n_\mathrm{QL}(0,t;x_0) = \lim_{x\rightarrow\infty}n_\mathrm{QL}(x,t;x_0) = 0$. 
 
There are two corrections to the diffusion model of \citet{yabushita_exact_1980} that we apply 
in order to capture the statistical evolution of test particles using the Fokker--Planck model in Equation \eqref{eq:FP_eqn_simpler}.
First, the existence of a last KAM curve at some sufficiently large energy, $x_\mathrm{KAM}$, implies that a more appropriate boundary condition when solving Equation \eqref{eq:FP_eqn_simpler} is to impose a reflecting boundary satisfying  $\partial_x\left(x^{3/2}n(x,t)\right)_{x=x_\mathrm{KAM}}=0$. 
Second, as we showed in Section \ref{sec:diffusion:local_diffusion}, the diffusion coefficient  $\mathcal{D}(\epsilon,\beta)$ can deviate significantly from the quasi-linear diffusion rate near this upper boundary.  Below in Section \ref{sec:diffusion:nbody_compare}, we will take these deviations into account by fitting a simple parametric form to the calculated local diffusion coefficient, $\mathcal{D}(\epsilon,\beta)$.

We now investigate the first of these corrections. Equation \eqref{eq:FP_eqn_simpler} subject to the boundary conditions $n(0,t)=\partial_x\paren{x^{3/2}n(x,t)}_{x=x_\mathrm{KAM}}=0$  can be cast as a Sturm-Liouville problem. By writing  $n(x,t)=x^{-3/2}\sum_{i}\phi_i(x)\exp\paren{-\frac{\lambda_i}{2T_D}t}$ and defining $p(x) = \mathcal{D}(\epsilon(x))/\mathcal{D}_\mathrm{QL}$, we  obtain the differential equation  
\begin{equation}
\label{eq:sturm_liouville}
    \dd{}{x}\left( p(x) \phi_i'(x)\right)+\lambda_i x^{-3/2}\phi_i(x) = 0~,
\end{equation}
for the functions $\phi_i$, subject to the boundary conditions $\phi_i(0)=0$ and $\phi_i'(x_\mathrm{KAM})=0$.
When $p(x) = 1$, i.e., when we approximate  $\mathcal{D}(\epsilon(x))$ as $\mathcal{D}_\mathrm{QL}$, Equation \eqref{eq:sturm_liouville} admits
the closed-form solutions
\begin{equation}
\label{eq:fp_eigenmodes}
    \phi_i(x) = 
    \frac{\sqrt{x} J_2(4 \sqrt{\lambda_i}{x}^{1/4})}{\sqrt{2}J_2(j_{1,i})x_\mathrm{KAM}^{1/4}}~
\end{equation}
 with eigenvalues 
\begin{equation}
\label{eq:fp_eigenvalues}
   \lambda_i = \frac{j_{1,i}^2}{16 \sqrt{x_\mathrm{KAM}}}
\end{equation}
where $j_{1,i}$ denotes the $i$th root of the Bessel function, $J_1$. The normalizations of the eigenfunctions in Equation \eqref{eq:fp_eigenmodes} have been chosen so that $\int_{0}^{x_\mathrm{KAM}}\phi_i(t)\phi_j(t)t^{-3/2}dt = \delta_{i,j}$. Given an initial density, $n(x,0) = \delta(x-x_0)$, the solution for the time-dependent density is
\begin{equation}
    n(x,t) = \sum_{i=1}^\infty
    x^{-3/2}
    \phi_i(x_0)
    \phi_i(x)
    \exp\left[-\frac{\lambda_i}{2T_D}t\right]~.
    \label{eq:fp_density}
\end{equation}
Using  $\int_0^{x_\mathrm{KAM}}\phi_i(t)t^{-3/2}dt = \sqrt{2}/({J_2(j_{1,i})x_\mathrm{KAM}^{1/4}})$, the fraction of particles, $f_\mathrm{survive}$, surviving up to a time, $t$, is obtained by integrating the expression for $n(x,t)$ in Equation \eqref{eq:fp_density}, yielding
\begin{equation}
\label{eq:fp_fsurvive}
f_\mathrm{survive}(t;x_0,x_\mathrm{KAM}) = \sum_{i=1}^\infty    \frac{\sqrt{x_0} J_2(4 \sqrt{\lambda_i}{x_0}^{1/4})}{|J_2(j_{1,i})|^2x_\mathrm{KAM}^{1/2}}
\exp\left[-\frac{\lambda_i}{2T_D}t\right]~.
\end{equation}
 Since\footnote{\href{https://dlmf.nist.gov/10.21}{NIST DLFM 10.21(vi)}} $j_{1,i} \approx \paren{i + \frac{1}{4}}\pi +\mathcal{O}{(1/i)}$, calculating the sums appearing in Equations \eqref{eq:fp_density} or \eqref{eq:fp_fsurvive} to a numerical accuracy of $\delta$ requires retaining terms up to order $i_\mathrm{max}\sim \sqrt{\frac{-T_D\log\delta}{t}}$.
While the sum in Equation \eqref{eq:fp_fsurvive} is apparently divergent at $t=0$, we confirm numerically that for $0< t/T_D\ll 1$, it converges to $\sim 1$ when a sufficient number of terms are included.
\subsection{Comparison to N-body integrations}
\label{sec:diffusion:nbody_compare}
\begin{figure}
    \centering
    \includegraphics[width=0.85\textwidth]{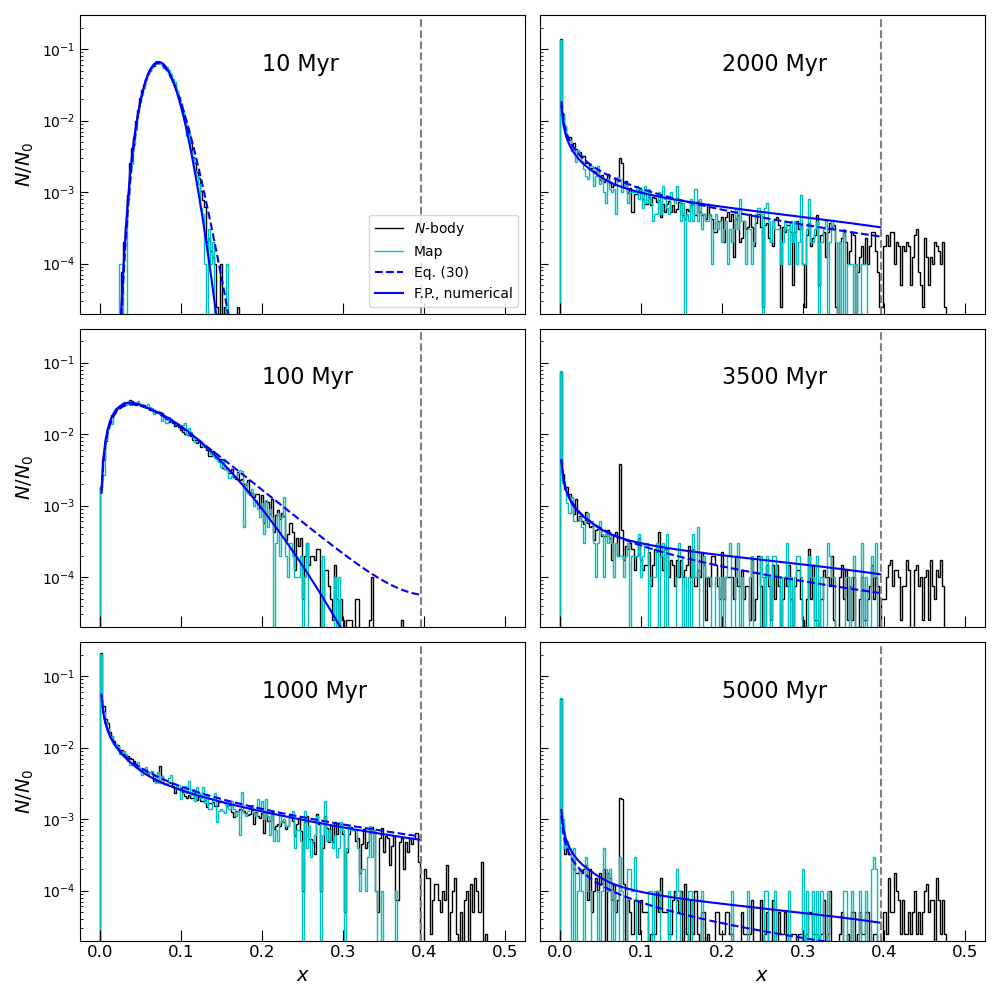}
    \caption{ 
    Comparison between particle orbital energy distributions at different times predicted by 
    direct $N$-body simulations,
    the iterated mapping,
    and the Fokker--Planck equation, for an ensemble of test particles subject to a Neptune-mass perturber at $a_N=30\AU$.
    Particles are initialized with $a = 400\AU$ and a pericenter distance of $q=35\AU$.
    Histograms in each panel show the number of particles, normalized by the initial number of particles, $N_0$, in bins of $x=a_N/a$.
    The predictions of the Fokker--Planck models are plotted as blue lines. 
    The dashed blue line shows the analytic prediction of Equation \eqref{eq:fp_density}, obtained by approximating the diffusion coefficient as $D(x) = 4\mu^2x^{3/2}\mathcal{D}_{QL}(a_N/q)$, while the solid blue line shows a numerical solution to the Fokker--Planck equation adopting  the diffusion coefficient given in Equation \eqref{eq:D_parameteric_fit}. 
    The Fokker--Planck equations are solved with a reflecting boundary at $x_{KAM} = 4^{-2/3}$, indicated by the vertical dashed line in each panel.}
    \label{fig:Nbody_vs_FP_histograms}
\end{figure}

\begin{figure}
    \centering
    \includegraphics[width=0.75\textwidth]{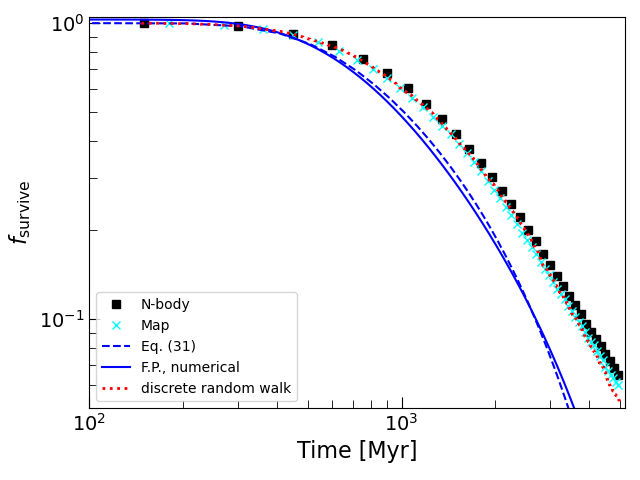}
    \caption{
    Fraction of surviving particles versus time for an  ensemble of particles having initial semimajor axis $a=400\AU$ and pericenter distance $q=35\AU$ and subject to a Neptune-mass perturber on a circular orbit at $30\AU$.
    The black squares show $N$-body results while cyan crosses indicate results from the mapping.
    The blue dashed line shows the prediction of the quasi-linear Fokker--Planck model, given by Equation \eqref{eq:fp_fsurvive}, while the solid blue curve shows the results of a numerical integration of the Fokker--Planck equation using the diffusion coefficient given by Equation \eqref{eq:D_parameteric_fit}.
    The red dotted line shows the results from the discrete random walk (Equation \ref{eq:newmap}). 
    }
    \label{fig:Nbody_vs_FP_fsurvive}
\end{figure}

We compare our Fokker--Planck model and the iterated mapping to direct $N$-body integrations in Figures \ref{fig:Nbody_vs_FP_histograms} and \ref{fig:Nbody_vs_FP_fsurvive}. The $N$-body integrations  follow an ensemble of $4\times 10^4$ test particles with initial pericenter distance $q=35\AU$ and semi-major axes $a = 400\AU$, perturbed by a Neptune-mass planet on a circular orbit at $30\AU$ about a $1 M_\odot$ star.
Integrations are done with the REBOUND \texttt{WHFast} integrator \citep{rein_whfast_2015} using a time step equal to 1/20th of Neptune's orbital period.
We also evolve an ensemble of $10^4$ map orbits, initialized with $x=30/400$ and random initial $\theta$ values. For these orbits, we augment the mapping Equations \eqref{eq:map_full1} and \eqref{eq:map_full2} with an auxillary time variable, $t$, updated at each step according to  $t' = t + P_Nx^{-3/2}$. 

Figure \ref{fig:Nbody_vs_FP_histograms} shows the normalized histograms of test particle energies, $x$, for $N$-body integrations and map orbits at different times. 
The histograms show mostly excellent agreement, except that a handful of $N$-body particles reach $x$ values between the 4:1 MMR (indicated by the dashed vertical line) and the 3:1 MMR, whereas the map orbits are excluded from this region.
The disagreement between the mapping and $N$-body integrations regarding the behaviour near the 4:1 MMR can be attributed to the mapping model's assumption of a constant pericenter distance.  Conservation of the Tisserand parameter implies that particles initialized 
with a pericenter distance of $q = 35\AU$ and $a=400\AU$ will have a slightly smaller pericenter distance of $q \approx 34.5\AU$ upon reaching the 4:1 MMR.
We find by conducting the test to determine the last KAM barrier described in Section \ref{sec:chaos}, that at this slightly lower pericenter distance, there is no longer a bounding KAM at the 4:1 MMR, though the flux of particles across the resonance is extremely slow. (It requires on the order of $\sim 5 \times 10^4$ map iterations, corresponding to $\sim30\mathrm{~Myr}$, for the first particle from an ensemble of $10^3$ particles initialized in the vicinity of $(\theta,w) = 0.5$ to escape the region $0<w<1$.)
No $N$-body particles reach period ratios interior to the 3:1 MMR, consistent with  additional numerical tests  with the mapping that indicate a bounding KAM curve exists near this MMR when $q = 34.5\AU$.

Figure  \ref{fig:Nbody_vs_FP_histograms} also compares the predictions of two Fokker--Planck models to the particle energy distributions obtained with the mapping and $N$-body simulations.
For both models, we impose an absorbing boundary at $x=0$ and a reflecting boundary at $x_\mathrm{KAM} = 4^{-2/3}$.
The first Fokker--Planck model, which we refer to as the ``quasi-linear'' model, approximates the diffusion coefficient as $D(x) = 4 \mu x^{3/2}\mathcal{D}_\mathrm{QL}(q/a_N)$ and the corresponding density distribution is predicted by Equation \eqref{eq:fp_density}.
For the second model, which we will refer to as the ``numerical'' model, we adopt a simple parametric model for the diffusion coefficient given by
\begin{equation}
    D(x) = 4 \mu^2x^{3/2}\mathcal{D}_{QL}\left[
    1+a_1 e^{-b \epsilon(x)}\log(\epsilon(x))
    + 
    a_2 \exp\left( 
    -\frac{(\epsilon(x) -\epsilon_*)^2}
    {2 \Delta_*^2}
    \right)
    \right],
    \label{eq:D_parameteric_fit}
\end{equation} 
where $\epsilon(x) = 3\mu x^{-5/2}$ and the parameter values $(a_1,b,a_2,\epsilon_*,\Delta_*) = (0.136573, 0.971338, 0.286504, 0.269842, 0.170774)$ are obtained from a fit to the data shown in Figure \ref{fig:diffusion_compare}.
Equation  \eqref{eq:FP_eqn_simpler} is then solved numerically to obtain the density distribution, $n(x,t)$, at the times shown in Figure \ref{fig:Nbody_vs_FP_histograms}.
The consequences of the reduced diffusion rate at values of $x\gtrsim 0.2$ are especially evident at 100 Myr, where the numerical model predicts a much smaller fraction of particles at $x\gtrsim 0.2$ than the quasi-linear model.
The numerical model also predicts a significantly larger fraction of particles surviving in this region at late times compared to the quasi-linear model.
Both Fokker--Planck models significantly under-predict the number of very loosely bound particles evident in the smallest histogram bins.
In this region, the approximation of particle orbits as a time-continuous random process, which underlies the derivation of the Fokker--Planck equation, breaks down, as we now show. 

Figure \ref{fig:Nbody_vs_FP_fsurvive} plots the fraction of surviving particles versus time for both the mapping and $N$-body simulations and compares these to the prediction of Equation \eqref{eq:fp_fsurvive}.
The mapping and $N$-body integrations show good agreement, while the Fokker--Planck model severely under-predicts the fraction of bound particles at late times.
We find that the disagreement is due to the fact that at $x\ll 1$ the approximation of trajectories as a time-continuous random walk breaks down.  To demonstrate this we simulate particle trajectories, $\left\{(t_1,x_1),(t_2,x_2),...\right\}$, as a discrete random walk, where $(t_1,x_1) = (P_N(a/a_N)^{3/2},a_N/a)$ and 
\begin{align}
    t_{i+1} &= t_i + P_N x_i^{-3/2}\\
    x_{i+1} &= x_\mathrm{KAM} - | x_i + 2\mu \sqrt{\mathcal{D}_{QL}(q/a_N)} \Delta_i - x_\mathrm{KAM} |
    \label{eq:newmap}
\end{align}
with $\Delta_i \sim \mathcal{N}(0,1) $  a zero-mean Gaussian random variable with unit variance.
The random walks are stopped at the first step, $i$, for which $x_{i}<0$.
We simulate $2^{17}$ random walks and plot the fraction of surviving particles as a function of time in Figure \ref{fig:Nbody_vs_FP_fsurvive}. 
The discrete random walk shows excellent agreement with the survival fractions of the  mapping and $N$-body simulations.

\subsection{Resonance sticking}

\label{sec:sticking}

The phenomenon of ``resonance sticking'' can cause the evolution of chaotic trajectories to deviate from a random-walk process and lead to power-law tails in the distribution of times needed for trajectories to escape from regions of phase space containing a mixture of regular and chaotic trajectories  \citep[e.g.,][]{meiss_thirty_2015}.
In their study of evolution of comets on Neptune-crossing orbits, \citet{malyshkin_keplerian_1999} find that the population of comets surviving up to $\sim 5$ Gyr is dominated by orbits that experience prolonged periods of resonance sticking.
By contrast, the agreement between the survival times of the discrete random-walk model, which cannot produce sticking, and the $N$-body integrations presented in Figure \ref{fig:Nbody_vs_FP_fsurvive} indicates that resonance sticking plays a minimal role in the survival of particles up to 5 Gyr for the initial semi-major axis and pericenter distance adopted in these experiments.
The influence of sticking is not apparent in the simulations of \citet{malyshkin_keplerian_1999} until the comets are depleted by a factor of $\sim 10^{-4}$, whereas the total particle population in Figure \ref{fig:Nbody_vs_FP_fsurvive} is only depleted by a factor of $\sim 0.05$ after 5 Gyr, so it is not entirely surprising that sticking has a negligible influence on our survival statistics.
Nonetheless, we find that resonance sticking does occur in our map and  we briefly explore its effect on the escape times from local regions of phase space below.

\begin{figure}
    \centering
    \includegraphics[width=0.95\textwidth]{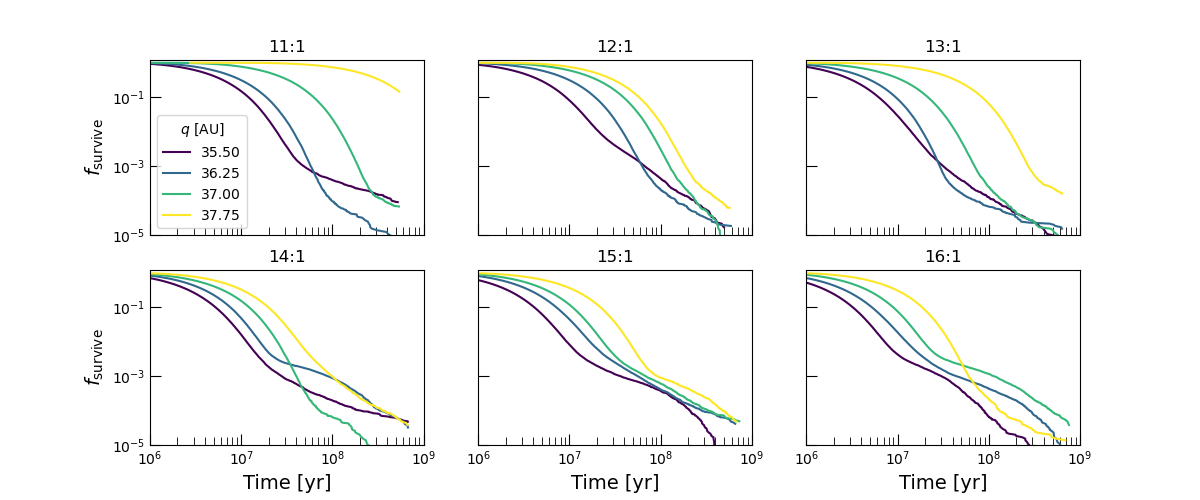}
    \caption{
    Survival fraction versus time for chaotic orbits escaping the vicinity of $N$:1 MMRs of Neptune at different pericenter distances. The fraction of orbits, $f_\mathrm{survive}$, from an initial ensemble of $655,360$ orbits, that remain within $-1<w<1$ is plotted against time (computed as $N P_N$ times the number of iterations). 
    Different color curves show different pericenter distances. 
    }
    \label{fig:escape_time_dist}
\end{figure}

In Figure \ref{fig:escape_time_dist}, we use the local map to compute the fraction of orbits remaining in the vicinity of $N$:1 MMRs with $11\le N \le 16$ as a function of time for a range of pericenter distances.
Here and throughout this section, all map parameters are reported assuming a Neptune-mass perturber located at an orbital distance of $a_p = 30\AU$.
For each $N$:1 MMR and pericenter distance, we initialize an ensemble of 655,360 orbits near the unstable fixed point of the first-order resonance at $(\theta,w) = (0,0)$ by drawing their initial $w$ values from a Gaussian centered at $w=0$ with dispersion $\sigma_w = 10^{-5}$ while setting their initial $\theta=0$. 
Each orbit is then evolved until it escapes the region $-1<w<1$.
The fraction of surviving orbits, $f_\mathrm{survive}$, is plotted against time in Figure \ref{fig:escape_time_dist} (here time is taken to be the number of map iterations times $N\times P_N$ where $N$ is set by the particular MMR).

\begin{figure}
    \centering
    \includegraphics[width=0.75\textwidth]{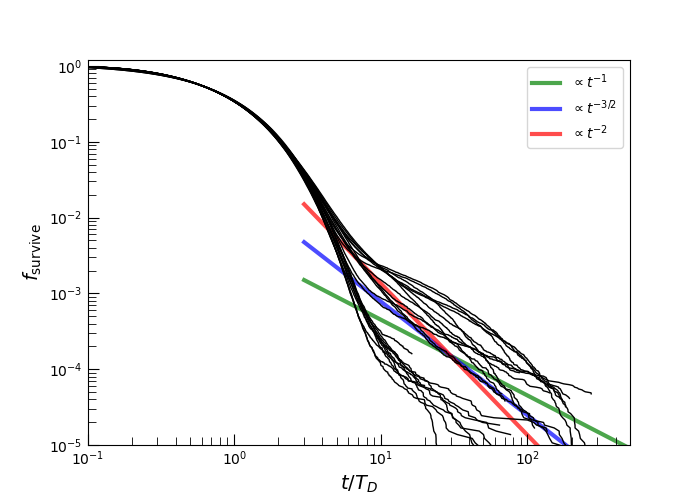}
    \caption{
    Survival fraction versus time normalized by a best-fit decay time scale, $T_D$.
    Each curve shows the survival fraction for one of the parameter combinations of $N$ and pericenter distance, $q$, plotted in Figure \ref{fig:escape_time_dist}.  
    Power-law slopes of $-1,-3/2$ and $-2$ are indicated by colored lines for comparison.
    }

    \label{fig:normalized_escape_times}
\end{figure}

For $f_\mathrm{survive}\gtrsim 0.01$, all of the curves in Figure  \ref{fig:escape_time_dist} exhibit the exponentially decaying behaviour expected for a normal diffusive process.
This is illustrated clearly in Figure \ref{fig:normalized_escape_times}, where we re-plot the survival fraction curves shown in Figure \ref{fig:escape_time_dist}, but with the time coordinate re-scaled by a best-fit exponential decay time scale, $T_D$. 
(We determine the decay time scale, $T_D$, by means of a least-squares fit of each $f_\mathrm{survive}(t)$ curve shown in Figure \ref{fig:escape_time_dist} using a simple exponential decay model, $e^{-t/T_{D}}$.)
For $f_\mathrm{survive}\lesssim 0.01$, many of the curves exhibit apparent algebraic decays at late times, indicative of resonance sticking \citep[e.g.,][]{karney_long-time_1983}.
As is evident in Figure \ref{fig:normalized_escape_times}, there does not appear to be a universal power-law slope for the decay of survival fractions, which instead ranges approximately between $t^{-2}$ and $t^{-1}$ behaviour.

\begin{figure}
    \centering
    \includegraphics[width=0.95\textwidth]{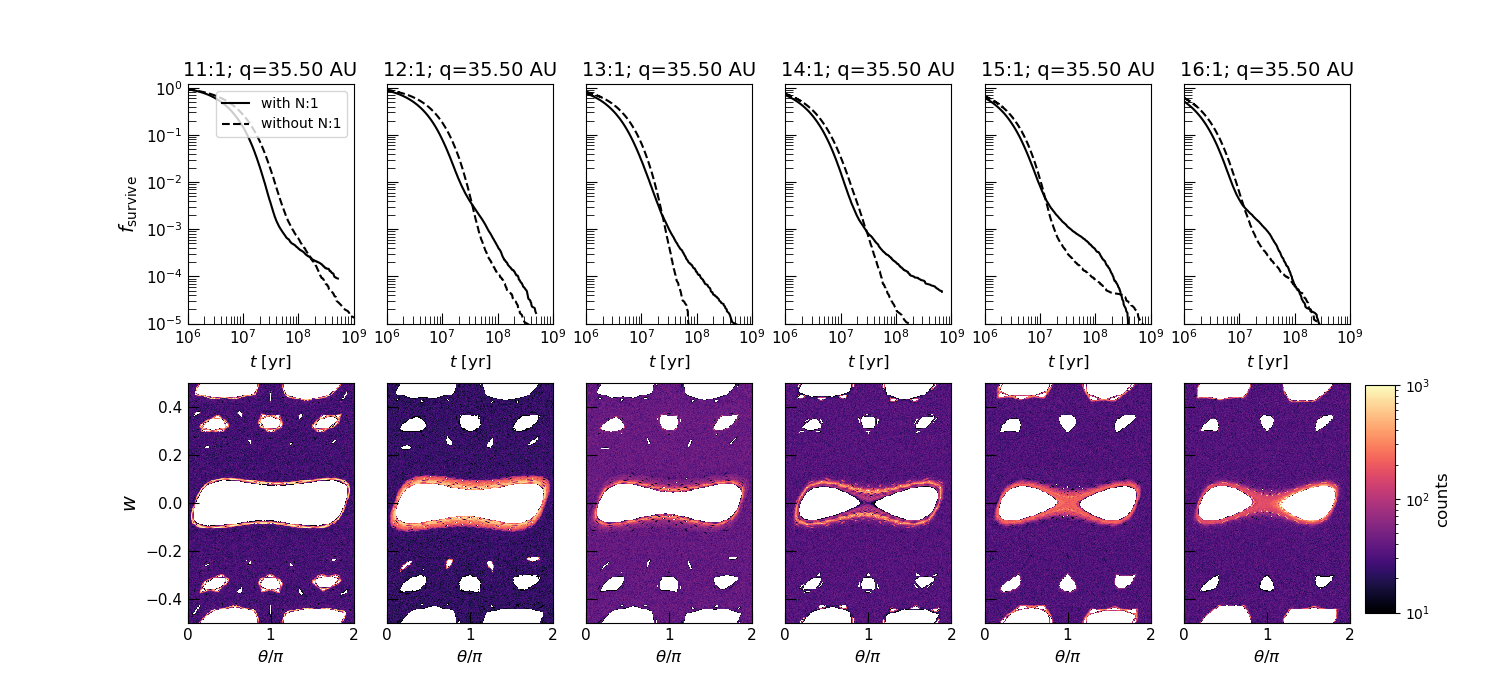}
    \caption{
    Top panels show the surviving fraction of orbits in the neighborhoods of various $N$:1 MMRs as a function of time. 
    To highlight the contribution of resonance sticking in $N$:1 MMRs, the solid lines show the surviving fraction of orbits initialized near $w=0$  that remain in the region $-1 < w < 1$, while the dashed lines show the surviving fraction of orbits initialized near $w=0.5$  that remain in the region $0 < w < 1$. (Exit time values for the latter curves are multiplied by a factor of 4 in order to account for the smaller region size.)
    The bottom panels show two-dimensional histograms constructed from the 100,000 iterations 
    of the $0.1\%$ of trajectories from an initial ensemble of $32,768$ initialized near $w=0$ that remain in the region $-1 < w < 1$ the longest. The $w$ values of these trajectories are plotted modulo 1 in the histograms.
    }
    \label{fig:sticking_example_q35}
\end{figure}

Closer inspection reveals that changes in the late-time survival fraction behaviour at a given pericenter distance are associated with
changes in the phase-space topology of the $N$:1 MMRs as $N$ varies. 
This is illustrated in Figure \ref{fig:sticking_example_q35},
where we plot a series of survival fraction curves for $q=35.5\AU$ at different $N$:1 MMRs accompanied by two-dimensional histograms  showing the densities of the  trajectory points of the longest-surviving orbits.
Specifically, the histograms are
constructed by recording the trajectories of 32,768 map points initialized near $(\theta,w) = (0,0)$ and binning the trajectory points of the 
33 orbits ($=0.1\%$ of $32,768$) that remain in the region $-1<w<1$ the longest. 
%
To highlight the contribution of sticking in $N$:1 MMRs to the survival fraction behaviour, we compare survival fractions versus time in two different regions near each $N$:1 resonance in the top panels of Figures \ref{fig:sticking_example_q35}.
The first region we consider includes the range $-1<w<1$, as in Figure \ref{fig:escape_time_dist}, with trajectories initialized near the $N$:$1$ MMR located at $w=0$.
The second region includes only the range $0<w<1$, and we initialize trajectories near the  $(2N+1)$:$2$ MMR, with points near $(\theta,w) = (0,1/2)$. Since neighboring $N$:1 MMRs sit on the boundary of this second region, orbits that stick to the boundary of these MMRs quickly exit the region $0<w<1$, and so sticking to $N$:1 MMRs has negligible influence on the escape times for this second region.
The resulting survival fractions for the first and second regions are plotted as solid and dashed curves, respectively, in Figure \ref{fig:sticking_example_q35}. Exit time values for the second region are multiplied by a factor of 4 to account for the smaller region size.\footnote{If the orbits' evolution were described perfectly by time-continuous diffusion processes with a spatially constant diffusion coefficient, then their escape times would be invariant under a simultaneous re-scaling of length by a factor of $l$ and time by a factor of $l^2$. We find that differences between the dashed and solid curves in Figure \ref{fig:sticking_example_q35} at values of $f_\mathrm{survive}\gtrsim 10^{-3}$, where the influence of sticking is negligible, are apparently due to variations of the diffusion rate rather than inaccuracies of the continuous-time approximation. We simulated two ensembles of discrete random walks initialized at $w=0$ and $w=1/2$, and with diffusion rates set to the
rate measured numerically following the procedure described in Section \ref{sec:diffusion:local_diffusion}, taking $N=16$  and $\beta = 30/35.5$. We recorded the escape times of the two ensembles from the regions $-1<w<1$ and $0<w<1$, respectively and found them to be nearly identical after multiplying the escape times of the second ensemble by a factor of 4.
}

Examining the escape time distributions in Figure \ref{fig:sticking_example_q35} simultaneously with the phase space density of the longest-surviving orbits, it is clear that $N$:1 MMRs produce the most dramatic sticking effects. 
Furthermore,  it appears that 
the $N$:1 MMRs are ``stickiest'' at or very near the value of $N$ at which chaotic orbits can first traverse the region between the two asymmetric libration islands on either side of the hyperbolic fixed point at  $(\theta,w) = (\pi,0)$. 
We observe similar behaviour across all  pericenter distances in our numerical experiments: 
sticking in $N$:1 MMRs is most significant near the values of $N$ at which the last invariant symmetric libration curve (i.e., the last regular $N$:1 resonant trajectory in which the resonance angle librates symmetrically about $\pi$) vanishes.

\section{Summary and Conclusions}
\label{sec:summary_and_conclusions}

We have presented a new iterated map for modeling the dynamical evolution of highly eccentric test particles subject to a planetary mass perturber. 
Utilizing a new formula for the energy kick received by test particles at each perihelion passage, 
the map extends the validity of previous mapping treatments of this problem to semi-major axes and pericenter distances relevant to the bulk of the observed extreme TNO population. 
The map shows excellent agreement with direct $N$-body integrations, at a dramatically reduced computational cost, while also elucidating how the dynamics depend on perturber mass and orbital period through the local approximation introduced in Section \ref{sec:local_map}.

In Section \ref{sec:resonances}, we showed how the new kick function can be used to derive an integrable Hamiltonian model for the dynamics of distant mean-motion resonances (MMRs).
These models are well-suited to study the resonant dynamics of highly eccentric, distant TNOs and do not require resorting to the usual expansions in powers of semi-major axis ratio or eccentricities. 
Our Hamiltonian model for MMRs also allows us to predict resonance widths and to determine, for a given pericenter distance and perturber mass, the semi-major axis at which resonance overlap leads to the onset of dynamical chaos. 
We provide a novel analytic criterion for the onset of chaos, given by Equation \eqref{eq:tau_res}, that agrees well with numerical simulations.
This new criterion improves upon the resonance overlap criterion of \citet{batygin_stability_2021}, which approximates the onset of chaos at large semi-major axis but performs poorly at smaller semi-major axes more typical of the observed TNO population.

In Section \ref{sec:diffusion}, we have presented a statistical description of the transport of chaotic orbits. 
 We have shown how the mapping can be used to determine the local energy diffusion rate in Section \ref{sec:diffusion:local_diffusion}.
 We demonstrate that, in regions where regular islands occupy a non-negligible fraction of phase space, long-time correlations can significantly reduce the diffusion rate relative to a naive prediction that treats successive energy kicks during pericenter passage as uncorrelated.
Examining the diffusion rates shown in Figure \ref{fig:diffusion_compare} at semi-major axes and pericenter distances typical of the observed TNO population shown in Figure \ref{fig:overlap_prediction}, it is clear that the effects of correlations 
can play a significant role in the chaotic transport of much of this population.

The local diffusion rates derived from the map in Section \ref{sec:diffusion:local_diffusion} serve as the basis for a Fokker--Planck description of the global transport of particles in energy space, derived in Section \ref{sec:diffusion:FP}.
Our Fokker--Planck model reduces to the diffusion equation treated in previous works \citep[e.g.,][]{van_woerkom_origin_1948,shteins_diffusion_1961,yabushita_exact_1980} when one ignores deviations from the quasi-linear value of the diffusion coefficient introduced by correlations and approximates it as $D(x)\propto x^{3/2}$. 
Note that the mapping enables us to compute the value of the diffusion coefficient explicitly,
whereas past works have either relied on order-of-magnitude estimates or simply worked in re-scaled time units so that the diffusion constant is unity. 
Additionally, the existence of KAM tori at short orbital periods for a given pericenter distance (or, more precisely, Jacobi constant) results in a reflecting boundary condition at some orbital energy, $x_\mathrm{KAM}$, rather than the absorbing boundary at $x\to\infty$ used by \citet{yabushita_exact_1980}.
We derived a new analytic solution (Equation \ref{eq:fp_density}) to the Fokker--Planck problem under these boundary conditions, assuming the quasi-linear diffusion approximation.

We demonstrated in Section \ref{sec:diffusion:nbody_compare} that our simple Fokker--Planck formulation provides good agreement with direct $N$-body integrations. Our comparisons in Figure \ref{fig:Nbody_vs_FP_histograms} reveal that deviations of the diffusion rate from its quasi-linear value can have a large influence on the number of particles at small semi-major axes (i.e., large $x$), where they are most observable.
We also showed that the discrete nature of the random walk executed by particles, which maintain essentially constant orbital energies between successive pericenter passages, enhances the number of very loosely bound particles relative to the predictions of the Fokker--Planck model, which assumes a continuous-time random walk.

Finally, in Section \ref{sec:sticking}, we showed that sticking in MMRs leads to late-time algebraic decay in the fraction of particles remaining in local regions of phase space.
First-order $N$:1 MMRs generally produce the most significant sticking effects, especially near the orbital period at which the last regular symmetric libration trajectory vanishes for a given pericenter distance.
An interesting avenue of future investigation would be to explain the sticking behaviour around $N$:1 MMRs and the late-time algebraic decay of the survival-time distributions in terms of a Markov tree model for transport \citep[e.g.,][]{meiss_markov_1986,alus_statistics_2014,alus_universal_2017}, especially near critical parameter values at which the last invariant symmetric libration curve breaks up.

There are a number of prospects for refining the dynamical model we have developed here. 
First, we have only considered a circular and coplanar perturber. 
It is relatively straightforward to extend the perturbative treatment in Appendix \ref{appendix:kick_derivation} to derive the increments in inclination, ascending node, and argument of pericenter induced by an eccentric, inclined perturber with each perihelion passage. 
From these, it should be possible to derive higher-dimensional symplectic mappings to describe the orbital evolution of inclined, eccentric test particles subject to an eccentric, massive perturber. Extending the mapping approach to include inclination dynamics may be particularly important for modelling the chaotic transport of eTNOs: the $N$-body simulations of \citet[][]{gallardo_survey_2012} show that particles' chaotic semi-major axis diffusion can be intermittently interrupted by temporary capture into MMRs, during which time the inclination and pericenter distance can be raised by a Kozai-like resonance (see their Figure 14).
Kozai cycling while sticking to MMRs has also been proposed as a mechanism for stranding inclined, detached TNOs at high pericenter distances during the late stages Neptune's outward migration \citep{gomes_origin_2003}.
The map could also be augmented to capture the influence of additional interior perturbing planets, secular torques from the galactic tides, and stochastic perturbations from passing stars.
\section*{Acknowledgements}
We thank Brett Gladman for helpful discussions.
We thank Yukun Huang for helpful discussions and for sharing results of $N$-body simulations. 
S.H. acknowledges support by the Natural Sciences and Engineering Research Council of Canada (NSERC), funding references CITA 490888-16 and RGPIN-2020-03885.

\section*{Data Availability}

The codes used in this article will be shared on reasonable request.



\bibliographystyle{mnras}
\bibliography{references, extra_refs} 


\appendix

\section{Derivation of Kick Function}
\label{appendix:kick_derivation}
\subsection{Hamiltonian Formulation of Problem}
In order to formulate our mapping, we start from the equations of motion governing the evolution of a test particle orbiting a central star of mass $M_*$ and perturbed by a planet of mass $m_p$ on a circular orbit of radius $a_p$.
The test particle's evolution is governed by the Hamiltonian
\begin{equation}
    \label{eq:Hamiltonian}
    \mathcal{H}(M,\varpi,\Lambda,G,t) = -\frac{\mathcal{G}^2\mathcal{M}^2}{2\Lambda^2} + \mathcal{R}(M,\varpi,\Lambda,G,t)~,
\end{equation}
where $\mathcal{G}$ is the gravitational constant and
$\mathcal{M}=M_* + m_p$ is the combined mass of the star and perturbing planet. 
The canonical momenta are  $\Lambda = \sqrt{\mathcal{GM}a}$ and $G=\sqrt{\mathcal{GM}a(1-e^2)}$
with $a$ and $e$ the test particle's barycentric semi-major axis and eccentricity, respectively. 
The corresponding conjugate angle variables are  the particle's mean anomaly, $M$, and  the particle's longitude of pericenter, $\varpi$. 
The disturbing function, $\mathcal{R}$, in Equation \eqref{eq:Hamiltonian} is given by 
\begin{align}
    \mathcal{R}(\lambda,\varpi,\Lambda,G,t) &=\mathcal{GM}\left(\frac{1}{|\pmb{r}|}-\frac{\mu}{|\pmb{r} -(1-\mu)\pmb{r}_p|} -\frac{1-\mu}{|\pmb{r} +\mu\pmb{r}_p|}\right) \nonumber \\
     &=-\mathcal{GM}\mu\left(\frac{1}{|\pmb{r} -\pmb{r}_p|} - \frac{\pmb{r} \cdot \pmb{r}_p}{|\pmb{r}|^3} -\frac{1}{|\pmb{r}|}\right) + \mathcal{O}(\mu^2)
         \label{eq:disturbing_function}
\end{align}
where $\mu = m_p/\mathcal{M}$, $\pmb{r}$ is the position vector of the test particle, $\pmb{r}_p =  (a_p\cos(n_pt),a_p \sin(n_p t))$, with $n_p = \sqrt{\mathcal{GM}/a_p^3}$, is the star-planet separation vector, and $\pmb{r}\cdot\pmb{r}_p/|\pmb{r}|^3$ is the indirect term.

Next,  we introduce a new set of canonical variables that will prove convenient for studying  (nearly-)parabolic orbits.
We apply the time-dependent canonical transformation $(\Lambda,G,M,\varpi) \rightarrow (\mathcal{T},L,\tau,l)$ generated by\footnote{The canonical transformation $(q_i,p_i)\rightarrow (Q_i,P_i)$ generated by a type-3 generating function is defined by the equations $q_i=-\partial_{p_i}F_3(p,Q,t)$ and $P_j=-\partial_{Q_i}F_3(p,Q,t)$ and the transformed Hamiltonian, $\hat{H}(Q,P)$, is given by $\hat{H}(Q,P) = H(q(Q,P),p(Q,P)) + \partial_t F_3(p(Q,P),Q,t)$.} 
\begin{equation}
    F_3(\tau,l,\Lambda,G,t) = \left[G + \frac{(\mathcal{GM})^2}{2\Lambda^2n_p}\right]\tau -  ( l + n_p t) G~.
\end{equation}
The new canonical momenta are $\mathcal{T} = -(\mathcal{GM})^2/(2\Lambda^2n_p) - G$ and $L=G$ and the new Hamiltonian, $\hat{\mathcal{H}}$, no longer depends explicitly on time because the disturbing function depends on time and azimuth only through the combination $\varpi-n_pt=l-\tau$. The transformed Hamiltonian is given by 
\begin{equation}
    \hat{\mathcal{H}} = \mathcal{H} + \pd{F_3}{t} = n_p \mathcal{T} + \mathcal{R}(\mathcal{T},L,\tau,l)~.
\end{equation}
Note that $ -\mathcal{T} \propto \frac{1}{2}\mathcal{GM}/a + n_p\sqrt{\mathcal{GM}a(1-e^2)}$, i.e., it is proportional to the  Tisserand parameter.

The next step is to express the disturbing function, Equation \eqref{eq:disturbing_function}, explicitly in terms of the new canonical variables. 
For a parabolic orbit of pericenter distance $q$, the specific orbital angular momentum is $L =(2\mathcal{GM}q)^{1/2}$ and the position vector of the test particle can be written 
\begin{equation}
    \pmb{r} = \mathbf{R}(n_p t+l-\tau) \cdot \frac{L^2}{2 \mathcal{GM}}
    \begin{pmatrix}
    1-D(L,\tau)^2\\2D(L,\tau)
    \end{pmatrix},    
\end{equation}
where the auxiliary variable $D$ is related to the true anomaly, $f$, according to $D = \tan(f/2)$ and  $\mathbf{R}(n_p t+l-\tau)$ is the rotation matrix that rotates coordinates by the angle $n_p t+l-\tau$.
The variable $\tau$ satisfies
\begin{equation}
    \tau = 2^{1/2}\fracbrac{q}{a_p}^{3/2}\left[D(L,\tau) + \tfrac{1}{3}D(L,\tau)^3\right]
    \label{eq:tau_of_D}
\end{equation}
where $q =  L^2/(2 \mathcal{G}M)$ is the pericenter distance.
Using the identity $\sinh^3x = \frac{1}{4} \sinh (3 x)-\frac{3}{4}\sinh x$ and setting  $D = 2\sinh x$
to invert Equation \eqref{eq:tau_of_D} we can express $D$ explicitly in terms of $\tau$ as
\begin{equation}
    D(L,\tau) = 2 \sinh \left[\tfrac{1}{3} \sinh ^{-1}\left(\frac{3 \tau}{(2q/a_p)^{3/2}}\right)\right]~.
\end{equation}
\subsection{Derivation of Kick Function}
We want an  approximation to the change in orbital energy, $\delta E$, experienced by a particle during a (nearly) parabolic encounter, to first order in planet-star mass ratio $\mu$. 
Since $\mathcal{T}$ will be conserved before and after a parabolic encounter, we have that 
$\delta E = n_p\delta G = n_p\delta L$  so that
\begin{align}
     \delta E &= -n_p\pd{}{l}\int_{-\infty}^\infty \mathcal{R} dt +
     \mathcal{O}(\mu^2)=-2^{1/2}\fracbrac{q}{a_p}^{3/2} \pd{}{l}\int_{-\infty}^\infty \mathcal{R} \times(1+D^2)dD +\mathcal{O}(\mu^2)\\
     &:= \frac{\mathcal{G}m_p}{a_p}\pd{}{l}F_{a_p/q}(l) +\mathcal{O}(\mu^2)
\end{align}
with  
\begin{equation}
    F_\beta(\theta)= 
    \sqrt{\frac{2}{\beta}}
    \int_{-\infty}^\infty
    \left(\frac{1+x^2}{\Delta(x,\beta,\theta)}
    -
    1
    - 
    \frac{\beta\cos\theta}{(1+x^2)^2}
    \left\{
    (1-x^2)
    \cos\left[
        \Phi(x;\beta)
    \right]
    +
    2x
    \sin\left[
      \Phi(x;\beta)
    \right]
\right\}
\right)dx
\label{eq:F_integral}
\end{equation}
where
   \begin{align}       
    \Phi(x;\beta) &= \sqrt{\frac{2}{\beta^3}}\left(x+\frac{1}{3}x^3\right),\\
    \Delta(x,\beta,\theta)^2&={(1+x^2)^2 + \beta^2 
    - 2\beta(1-x^2)
    \cos\left[
        \theta - \Phi(x;\beta)
    \right]
    +
    4\beta x
    \sin\left[
        \theta - \Phi(x;\beta)
    \right]}.\label{eq:Delta_def}
\end{align}
The integral in Equation \eqref{eq:F_integral} can be evaluated numerically in order to determine the energy kick experienced by a test particle as a function of its pericenter distance, $q$, and the angle $l=\omega - n_p t + \tau$.
For $\beta>1$, substituting $x = \pm\sqrt{\beta-1}$ into Equation \eqref{eq:Delta_def} and solving for $\Delta = 0$ we find that the integrand is singular at 
\begin{equation}
    \theta_\mathrm{int}(\beta)= 
    -\tan ^{-1}
    \left(
    \frac{2 \sqrt{\beta -1}}
    {
     2-\beta
     }
    \right)
    +
    \frac{\sqrt{2} (\beta +2) \sqrt{\beta -1}}{3 \beta ^{3/2}}
    -
    \pi.\label{eq:theta_intersect}
\end{equation}

\subsection{Cosine series representation of kick function}
For non-crossing orbits the kick function, $F_\beta$, can be written as a cosine series 
\begin{equation}
    F_\beta(\theta) = \sum_{k=0}^\infty C_k(\beta)\cos(k\theta)~.
    \label{eq:F_beta_sum_defn}
\end{equation}
Explicit calculation of the cosine coefficients $C_k(\beta)$ will be useful for deriving analytic expressions for the resonance widths and for deriving an analytic expression for the onset of chaos. 
Using the definition of Laplace coefficients, 
\begin{equation}
     \frac{1}{\sqrt{1+\beta^2-2\beta\cos\phi}}= \frac{1}{2}\sum_{k=-\infty}^{\infty}b_{1/2}^{(k)}\left(\beta \right)\cos(k\phi),
\end{equation}
 for $\beta<1$, we can write the integrand of Equation \eqref{eq:F_integral} as 
\begin{equation}
      \frac{1}{2}
      \sqrt{
      \frac{2}{\beta}
      }
        \sum_{k = -\infty}^{\infty} 
            \left(
                b_{1/2}^{(k)}\left(\frac{\beta}{1+x^2}\right)
                -
                \delta_{k,1}\frac{2\beta}
                {1+x^2}
                -
                2\delta_{k,0}
            \right)\cos[k(\theta - \Phi(x;\beta) + 2\arctan(x))]
\end{equation}
so that the cosine amplitudes $C_k$  for $k>1$ are given by 
\begin{align}
    C_k(\beta) &= \sqrt{\frac{2}{\beta}}\int_{-\infty}^\infty 
        b_{1/2}^{(k)}\left(\frac{\beta}{1+x^2}\right) 
    \cos\left[\sqrt{\frac{2}{\beta^3}}k\left(x+\frac{1}{3}x^3\right) - 2k\arctan(x)\right]
    dx\label{eq:Ck_defn}
    \\
    &:=\int_{-\infty}^\infty dx\left\{
    c_k(x;\beta)\cos\left[\sqrt{\frac{2}{\beta^3}}k\left(x+\frac{1}{3}x^3\right)\right]
    + 
    s_k(x;\beta)\sin\left[\sqrt{\frac{2}{\beta^3}}k\left(x+\frac{1}{3}x^3\right)\right]\right\}
\end{align}
where 
\begin{align}
     s_k(x;\beta)&=
     \sqrt{\frac{2}{\beta}}
        {b}^{(k)}_{1/2}\left(\frac{\beta}{(1+x^2)}\right)
     U_{2k-1}\left(\frac{1}{\sqrt{1+x^2}}\right)\times\fracbrac{x }{\sqrt{1+x^2}}\\
     c_k(x;\beta)&=
     \sqrt{\frac{2}{\beta}}
        {b}^{(k)}_{1/2}\left(\frac{\beta}{(1+x^2)}\right)
     T_{2k}\left(\frac{1}{\sqrt{1+x^2}}\right)
\end{align}
and $T_k$ and $U_k$ are Chebyshev polynomials.
When $k=1$, one should add to Equation \eqref{eq:Ck_defn} the contribution from the indirect term, 
\begin{align}
    C_{1,\mathrm{ind}} 
    &=
    -\beta\sqrt{\frac{2}{\beta}}
        \int_{-\infty}^{\infty}
            \frac{
            (1-x^2)\cos\Phi(x;\beta) + 2x\sin\Phi(x;\beta)
            }{
            (1+x^2)^2
            }    
    =
    -\sqrt{{2}\beta}        
    \int_{-\infty}^{\infty}
        \frac{\exp\paren{i\sqrt{\frac{2}{\beta^3}}(x+x^3/3)}}{(1+ix)^2}dx
    \nonumber\\
    &=   
    -{2\pi}\paren{
    \frac{2^{5/6} }{\sqrt{\beta }}
    \text{Ai}\left(\frac{2^{1/3}}{\beta }\right)
    -
    2^{2/3} \text{Ai}'\left(\frac{2^{1/3}}{\beta }
    \right)
   }
\end{align}
where $\text{Ai}$ and $\text{Ai}'$ are the Airy function of the first kind and its derivative.

Finally, using $\frac{1}{2}b_{1/2}^{(0)}(\alpha) = \frac{2}{\pi}\mathbb{K}(\alpha^2)$ where $\mathbb{K}$ is the complete elliptic integral of the first kind, the $k=0$ term in the sum in Equation \eqref{eq:F_beta_sum_defn} is given by 
\begin{equation}
\label{eq:C0_integral_defn}
    C_0(\beta)=\sqrt{\frac{2}{\beta}}\int_{-\infty}^\infty \paren{\frac{2}{\pi}\mathbb{K}\paren{\frac{\beta^2}{(1+x^2)^2}}-1}dx~.
\end{equation}
Using Hamilton's equations, it is straightforward to show that, to first order in $\mu$,  the particle's periapse precesses by an amount $\delta \varpi(\beta) =  \mu\sqrt{2\beta^3}\partial_\beta F_\beta(\theta)$ each perihelion passage. Outside of any mean-motion resonances, the particle's average secular precession rate is therefore given by $\dot{\varpi} = (2\pi)^{-1}\mu n_p x^{3/2}\sqrt{2\beta^3}\partial_\beta C_0(\beta)$. Expanding the integrand of Equation \eqref{eq:C0_integral_defn}, this is equivalent to 
\begin{equation}
   \dot{\varpi} = (2\pi)^{-1}\mu n_p x^{3/2}\sqrt{2\beta^3}\pd{}{\beta} \left\{\sqrt{\frac{2}{\beta}}\int_{-\infty}^\infty dx \paren{\frac{\beta ^2}{4 \left(x^2+1\right)^2}+\frac{9 \beta ^4}{64 \left(x^2+1\right)^4} +\mathcal{O}(\beta^6)}\right\} 
   \approx \mu n_p \fracbrac{a_p}{a}^{3/2}\paren{\frac{3 \beta ^2}{16} + \frac{315 \beta ^4}{2048}} \label{eq:precession_rate_hex}~.
\end{equation}
Upon substituting\footnote{In order to approximate elliptical orbits by parabolic ones, we have taken the limit $e\rightarrow 1$ and $a\rightarrow\infty$ at fixed angular momentum, $G = \sqrt{\mathcal{GM}a(1-e^2)} = \sqrt{2\mathcal{GM}q}$. Therefore, $\beta = a_p/q \rightarrow \frac{2a_p}{a(1-e^2)}$ is the appropriate substitution when comparing to rates derived using the usual secular disturbing function approach.}  $\beta = \frac{2a_p}{a(1-e^2)}$ in Equation \eqref{eq:precession_rate_hex}, we recover the precession rate predicted using the hexadecapolar-order expansion of the classical disturbing function given in  \cite{vinson_secular_2018}. In principle, we could account for this secular precession in our mapping by  amending Equation \eqref{eq:map_full2} to read $\theta' = \theta + 2\pi x'^{-3/2}  - \mu\sqrt{2\beta^3}\partial_\beta C_0(\beta)$. However, this  precession is of little practical consequence for the map dynamics and we ignore it in the main text.

\subsection{Large k limit}
\begin{figure}
    \centering
    \includegraphics[width=0.7\textwidth]{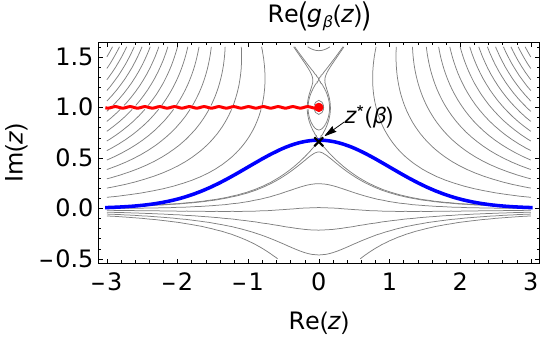}
    \caption{
     Contour plot showing the real part of the function $g_\beta(z)$ appearing in the integral defined by Eq.\  \eqref{eq:integral_k} for $\beta = 1/4$.
    The red line indicates a branch cut discontinuity.
    The contour $\mathcal{C}$ is indicated by the blue line, which passes through the saddle point, $z^*(\beta)$, indicated by a black `x', along the path of steepest descent. 
    }
    \label{fig:contour_int}
\end{figure}

For large values of $k$, Laplace coefficients are approximately given by \citep[][]{tremaine2023dynamics}
\begin{equation}
    b_{1/2}^{k}(\alpha) \approx \frac{2}{\sqrt{\pi | k| }}\frac{\alpha ^{| k|}}{  \sqrt{1-\alpha ^2}}~.
    \label{eq:laplace_approx}
\end{equation}
 Using
\begin{equation}
  \cos\left[\sqrt{\frac{2}{\beta^3}}k\left(x+\frac{1}{3}x^3\right) - 2k\arctan(x)\right] = \frac{1}{2}\exp\left[ik\sqrt{\frac{2}{\beta^3}}\left(x+\frac{1}{3}x^3\right) \right]\fracbrac{1 - i x}{1 + i x}^{k} + c.c.
\end{equation}
and Equation \eqref{eq:laplace_approx}, for $k>0$ we can write 
\begin{equation}
    C_k(\beta)
    \approx 
    \sqrt{\frac{2}{\beta}}  \frac{(-1)^k}{\sqrt{\pi k}}\beta^k
    \int_{-\infty}^\infty
    \frac{\left(1 + x^2\right)}{\sqrt{{\left(x^2+1\right)^2-\beta ^2}}}
     \frac{
        \exp
        \left[
            ik\sqrt{\frac{2}{\beta^3}}\left(x+\frac{1}{3}x^3\right)
            \right]
    }
    {
    (x - i)^{2k}
    } + c.c.
    \label{eq:large_k_integral}
\end{equation}
The integral on the right-hand side of Equation \eqref{eq:large_k_integral} can be approximated for large $k$ using the method of steepest descent.
To do so, define 
\begin{equation}
    \mathcal{I}_{\beta}(k)  = \int_\mathcal{C} f_\beta(z)e^{k g_\beta(z)}dz
    \label{eq:integral_k}
\end{equation}
where $f_\beta(z) = (1+z^2)((1+z^2)^2-\beta^2)^{-1/2}$,
$g_\beta(z) = i\sqrt{\frac{2}{\beta^3}}\left(z+\frac{1}{3}z^3\right) - 2\ln(z-i)$, and the contour $\mathcal{C}$ is shown in Figure \ref{fig:contour_int}.
Defining $z^*(\beta) = \frac{4}{3} i \sin \left(\frac{1}{3} \sin ^{-1}\left(1-\frac{27 \beta ^{3/2}}{8 \sqrt{2}}\right)\right)+\frac{i}{3}$, we have that $g'_\beta(z^*(\beta)) = 0$ and 
$\arg({g''(z^*)}) = \pi$ so the method of steepest descent gives the
 approximation of the integral defined in Equation \eqref{eq:integral_k} as 
\begin{equation}
    \mathcal{I}_{\beta}(k)  \approx f_\beta(z^*(\beta))e^{kg(z^*(\beta))}\sqrt{\frac{2\pi}{k|g''_\beta(z^*(\beta))|}}~.\label{eq:steepest_descent_result}
\end{equation}
Writing $z^*(\beta) = i \left(1-\fracbrac{\beta ^{3} }{{2}}^{1/4}R(\beta)\right)$
and combining Equations \eqref{eq:large_k_integral} and \eqref{eq:steepest_descent_result},
we arrive at the approximation
\begin{equation}
    C_k(\beta)\approx A(\beta)\frac{e^{- k \lambda(\beta) }}{k}
    \label{eq:asymp}
\end{equation}
where 
\begin{align*}
    \lambda(\beta) &=  
    \frac{2 \sqrt{2}}{3 \beta ^{3/2}}
    +
    \fracbrac{\beta^3}{2}^{1/4}\frac{ R^3}{3}
    -R^2+
    \frac{1}{2} \log \left(\frac{\beta  R^4}{2}\right)\\
    A(\beta) &= 
    \frac{
        2 {\beta^{1/4} } 
        R^2 
        \left(4-2^{3/4} \beta ^{3/4} R\right)
    }{
        \sqrt{
        \left(
             2+2 R^2+R^3-2^{3/4} \beta ^{3/4}
        \right)
        \left(
           4 \sqrt{2} R^2 -2 \sqrt{\beta }+\beta ^{3/2} R^4- 2^{9/4}\beta ^{3/4} R^3
        \right)
        }
    }~.
\end{align*}
In the limit $\beta\rightarrow 0$, we get $R\rightarrow 1$, $\lambda\rightarrow \frac{2\sqrt{2}}{3\beta^{3/2}} - 1 + \frac{1}{2}\log(\beta/2)$, and $A\rightarrow  2^{3/4}\beta^{1/4}$~.

The point, $z^*(\beta)$, at which the function $g_\beta$ exhibits a saddle occurs on the real axis for $\beta = 8/9$ so that the steepest descent method used to  approximate the integral in Equation \eqref{eq:integral_k} is no longer valid for $\beta \ge 8/9$. Interestingly, this value of $\beta$ is the limiting value, as $x\rightarrow 0$,  reached by particles with a Tisserand parameter of $T=3$ \citep[corresponding to an orbit that coincides with the planetary orbit; see, e.g.,][]{pan_generalization_2004}. In other words, particles with larger limiting values of $\beta$ can achieve orbits that cross the perturber's for some positive value of $x$.

\section{Derivation of Diffusion and Drift Coefficients}
\label{sec:diffusion_and_drift}
This appendix provides  details of derivations for diffusion and drift coefficients used in the statistical description of chaotic particles in Section \ref{sec:diffusion}.
\subsection{The diffusion/force correlation relationship}
\label{sec:diffusion-force}
Here we provide some of the intermediate steps in the derivation of Equation \eqref{eq:diff_coeff}.
The diffusion coefficient is defined as 
\begin{equation}
    \mathcal{D} = \lim_{n\rightarrow\infty}\frac{1}{\epsilon^2n}\langle (w_n - w_0)^2 \rangle,
\end{equation}
where $\avg{\cdot}$ denotes the ensemble average over phase space. Noting that $w_n = w_0 + \epsilon \sum_{k=0}^{n-1} a_k$,
\begin{eqnarray}
    \mathcal{D} = \lim_{n\rightarrow\infty}\frac{1}{n}\left\langle \paren{\sum_{k=0}^{n-1} a_k}^2\right\rangle
         = \lim_{n\rightarrow\infty}\frac{1}{n}\paren{\avg{\sum_{k=0}^{n-1} { a_k^2}} + 2 \avg{\sum_{r=0}^{n-1}\sum_{k = 1}^{r}a_{r-k} a_{r}}}
\end{eqnarray}
Stationarity implies
\begin{eqnarray}
\mathcal{D} = \avg{a_0^2} + 2\sum_{k = 1}^{\infty} \avg{a_{0}a_{k}}.
\end{eqnarray}
\subsection{Drift/diffusion coefficient relationship}
\label{sec:second_order_direct_derivation}
The goal of this Appendix is to derive the relation (\ref{eq:fluc-diss}) between the drift and diffusion coefficients in the comet map directly from Hamilton's equations.  

Write the Hamiltonian in action-angle variables as
\begin{equation}
H=H_0(J) + h(J,\phi,t),
\end{equation}
where $|h/H|$ is of order $\epsilon \ll 1$. The unperturbed trajectory, under the influence of $H_0(J)$ only, is $J=J_0$, $\phi=\phi_0+\Omega_0t$ with $\Omega_0=\partial H_0/\partial J|_{J_0}.$ The first-order perturbation to the action is obtained by integrating Hamilton's equations along the unperturbed trajectory,
\begin{equation}
\Delta_1J(t)=-\int_{-\infty}^t dt'h_\phi(t'),
\end{equation}
where $h_\phi(t')\equiv \partial h(J,\phi,t')/\partial\phi|_{J_0,\phi_0+\Omega_0t'}$.
Denote the angle average (i.e., average over $\phi_0$) by $\langle\cdot\rangle$. Then
\begin{equation}
\langle(\Delta_1J)^2\rangle=\Big\langle\Big[\int_{-\infty}^t dt'h_\phi(t')\Big]^2\Big\rangle.
\end{equation}
The diffusion coefficient in action is
\begin{equation}
D_J=\frac{d}{dt}\langle(\Delta_1J)^2\rangle=2\Big\langle h_\phi(t)\int_{-\infty}^t dt'h_\phi(t')\Big\rangle.
\end{equation}
Now differentiate with respect to $J$. Note that since $\Omega_0=H_J$ we have $dh_\phi(t)/dJ=(d/dJ)h_\phi(J,\phi+\Omega_0t,t)=h_{\phi J}(t) + H_{JJ}h_{\phi\phi}t$. We have
\begin{equation}
\frac{d D_J}{dJ}=2\Big\langle \big[h_{\phi J}(t) + H_{JJ}h_{\phi\phi}(t)t\big]\int_{-\infty}^t dt'h_\phi(t')+
h_\phi(t)\int_{-\infty}^t dt'\big[h_{\phi J}(t') + H_{JJ}h_{\phi\phi}(t')t'\big]\Big\rangle.
\label{eq:st1}
\end{equation}

The second-order perturbation to the action is
\begin{align}
\Delta_2J(t)&=-\int_{-\infty}^t dt'h_{\phi\phi}(t')\Delta_1\phi(t') - \int_{-\infty}^t dt'h_{\phi J}(t')\Delta_1J(t') \nonumber\\
&=-\int_{-\infty}^t dt'h_{\phi\phi}(t')\Big[H_{JJ}\int_{-\infty}^{t'} dt''\Delta J_1(t'') + \int_{-\infty}^{t'} dt''h_J(t'')\Big] + \int_{-\infty}^t dt'h_{\phi J}(t')\int_{-\infty}^{t'}dt''h_\phi(t'')\nonumber \\
&=\int_{-\infty}^t dt'h_{\phi\phi}(t')\Big[H_{JJ}\int_{-\infty}^{t'} dt''\int_{-\infty}^{t''}dt'''h_\phi(t''') - \int_{-\infty}^{t'} dt''h_J(t'')\Big] + \int_{-\infty}^t dt'h_{\phi J}(t')\int_{-\infty}^{t'}dt''h_\phi(t'').
\end{align}
Therefore the orbit-averaged drift coefficient in the action is
\begin{align}
B_J=\frac{d}{dt}\langle \Delta J_2\rangle 
&=\Big\langle h_{\phi\phi}(t)\Big[H_{JJ}\int_{-\infty}^{t} dt'\int_{-\infty}^{t'}dt''h_\phi(t'') - \int_{-\infty}^{t} dt'h_J(t')\Big] + h_{\phi J}(t)\int_{-\infty}^{t}dt'h_\phi(t')\Big\rangle.
\end{align}
Switching the order of the $t'$ and $t''$ integrations, and switching the labels of the dummy indices, 
\begin{align}
B_J
&=\Big\langle h_{\phi\phi}(t)\Big[H_{JJ}\int_{-\infty}^{t} dt'h_\phi(t')(t-t') - \int_{-\infty}^{t} dt'h_J(t')\Big] + h_{\phi J}(t)\int_{-\infty}^{t}dt'h_\phi(t')\Big\rangle.
\end{align}
Using Equation (\ref{eq:st1}),
\begin{equation}
\tfrac{1}{2}\frac{dD_J}{dJ}-B_J=\Big\langle h_\phi(t)\int_{-\infty}^t dt'h_{\phi J}(t')+ H_{JJ}h_\phi(t)\int_{-\infty}^t dt'h_{\phi\phi}(t')t' + H_{JJ}h_{\phi\phi}(t)\int_{-\infty}^t dt'h_\phi(t')t' + h_{\phi\phi}(t)\int_{-\infty}^t dt'h_J(t')\Big\rangle.
\end{equation}
Since we are orbit-averaging, we can integrate by parts over $\phi$. Then the first and fourth term cancel, as do the second and third. This proves that
\begin{equation}
    B_J=\tfrac{1}{2}\frac{d D_J}{d J}.
    \label{eq:bdrel}
\end{equation}
The drift and diffusion coefficients $B_J$ and $D_J$ are the mean and mean-square changes in the action per unit time. If the action is related to some other variable $x$ through $J=f(x)$ then the mean and mean-square changes in $x$, $B$ and $D$, are given by
\begin{equation}
    B_J=f'(x)B + \tfrac{1}{2}f''(x)D, \quad D_J^2=[f'(x)]^2D,
    \end{equation}
so Equation (\ref{eq:bdrel}) becomes
\begin{equation}
     B=\frac{1}{2f'}\frac{d }{d x}(f'D).
\end{equation}
If $x$ is the inverse semi-major axis then $f(x)\propto x^{-1/2}$ so $f'(x)\propto x^{-3/2}$ and we recover Equation (\ref{eq:fluc-diss}).


\bsp	
\label{lastpage}
\end{document}